# Bare-Die Antiferromagnetic Computing

*Yu Liu[#], Zhuoting Han[#], Zexin Feng[#], Peixin Qin*, Zhiyuan Duan, Yuhao Ye, Zengwei Zhu, Chengyan Zhong, Li Liu, Guojian Zhao, Wenbin Shen, Jingyu Li, Sixu Jiang, Xiaoyang Tan, Xiaoning Wang, Ziang Meng*, Chengbao Jiang*, Zhiqi Liu**

Y. Liu, P. Qin, Z. Duan, L. Liu, G. Zhao, J. Li, S. Jiang, X. Tan, Z. Meng, C. Jiang, Z. Liu
School of Materials Science and Engineering, Beihang University, Beijing 100191, China.
E-mail: qinpeixin@buaa.edu.cn; mengza@buaa.edu.cn; jiangcb@buaa.edu.cn; zhiqi@buaa.edu.cn

Y. Liu, P. Qin, Z. Duan, L. Liu, G. Zhao, J. Li, S. Jiang, X. Tan, Z. Meng, C. Jiang, Z. Liu
State Key Laboratory of Tropic Ocean Engineering Materials and Materials Evaluation, Beihang University, Beijing 100191, China.

Y. Liu, Z. Han, C. Zhong, W. Shen
College of Integrated Circuit Science and Engineering, Nanjing University of Posts and Telecommunications, Nanjing 210023, China.

Z. Feng
School of Electrical and Electronic Engineering, Nanyang Technological University, Singapore 637371, Singapore.

Y. Ye, Z. Zhu
Wuhan National High Magnetic Field Center, Huazhong University of Science and Technology, Wuhan 430074, China.

X. Wang
The Analysis & Testing Center, Beihang University, Beijing 100191, China.

[#]These authors contributed equally to this work.

**Abstract**
Semiconductor electronic devices are increasingly constrained by fundamental quantum tunneling effects and charge-based mechanisms, which severely limit further miniaturization, write-speed scaling, and environmental robustness of silicon-based technologies. These limitations are particularly prohibitive for deep-space exploration, where extreme temperatures, ultra-strong magnetic fields, and intense radiation rapidly incapacitate conventional electronics without massive shielding. Here, we present an intrinsically resilient, strain-mediated antiferromagnetic MnIr/PMN-PT edge processor that operates reliably as a bare die under temperatures up to 500 K, magnetic fields of 55 T, and radiation doses of 1.5 Mrad. By exploiting an input-modulated in situ self-refreshing encoding mechanism, the device performs nonlinear feature extraction and classification directly from raw analog signals, enabling an analog computing architecture that requires no time-frequency transformation. This architecture achieves 99.8% accuracy in speech recognition without digital preprocessing and 100% accuracy in astronaut visual object recognition. Furthermore, an all-hardware integrated drone vision system demonstrates real-time in situ command execution and autonomous navigation, delivering a terahertz-level response frequency and an ultra-low energy consumption of approximately 0.2 fJ per operation. This work expands the functional scope of antiferromagnetic devices beyond memory and logic, establishing them as a promising materials platform for energy-efficient physical computing and autonomous intelligence in extreme environments.

## 1. Introduction

As humanity's exploratory reach extends to the Jovian system, the Martian surface, and the cores of controlled fusion reactors, data processing in extreme physical environments faces unprecedented challenges.[1–13] In these scenarios, detectors must not only sense environmental parameters but also process massive analog signals and execute critical decisions within millisecond timescales. However, the enormous latencies imposed by communication distances (exceeding 40 minutes for Mars missions), as well as bandwidth constraints caused by intense electromagnetic interference render Earth-based data processing impractical.[14–16] This urgently demands computational hardware with *in-situ* computing capabilities, the ability to operate directly in hostile environments while independently completing a closed loop from signal acquisition and feature extraction to decision-making and control. Establishing such edge intelligence sovereignty is therefore paramount for endowing detectors with survival intelligence and rapid response capabilities in extreme conditions.

Yet, the prevailing computational paradigm faces a foundational crisis in extreme environments. Conventional charge-based microelectronics undergo systemic collapse without extensive shielding when exposed to high temperatures, intense magnetic fields, and ionizing radiation encountered in tokamaks, planetary magnetospheres, and particle accelerators.[17–20] The digital architecture further compounds this vulnerability: mandatory Analog to Digital Converter (ADC) chains and discrete logic states act as fragility multipliers, as charge carriers across all components succumb to radiation-induced disruption and thermal agitation.[21] While current engineering standards attempt to mitigate these failures through extensive shielding, this imposes prohibitive mass penalties (>$ 50,000 per kg) and inevitably degrading over time due to micrometeorite damage.[22–24] Crucially, such engineering workarounds fail in forbidden zones such as tokamak first-walls, where environmental fluxes exceed practical limits. Beyond these environmental constraints, semiconductor electronic devices rely on the electron's charge degree of freedom, for which quantum tunneling has become a fundamental bottleneck that severely constrains further miniaturization and write-speed scaling of silicon-based technologies in the future.[25–27] Mission survival in these regimes can no longer rely on protective engineering or computational sophistication but instead demands a revolutionary paradigm shift toward intrinsically resilient, physics-driven computing substrates.

Spintronic approaches offer a promising route beyond these post-Moore limitations by exploiting the electron's spin degree of freedom rather than its charge. However, spintronic devices based on ferromagnets remain fundamentally limited by their sensitivity to external magnetic fields, rendering them inoperable in the multi-tesla environments of interest.[28–30] Consequently, antiferromagnetic materials emerge as a viable physical platform, offering immunity to magnetic interference and stability at high temperatures, while also exhibiting ultrafast spin dynamics at terahertz frequencies, which surpasses the gigahertz-scale dynamics of conventional ferromagnetic materials.[31,32] However, a final bottleneck remains: conventional manipulation of antiferromagnetic order typically relies on high-density currents that induce substantial Joule heating, rendering such approaches unsuitable for power-constrained edge scenarios.[33,34]

In this work, we surmount this energy barrier by utilizing electric-field-induced piezoelectric strain to modulate antiferromagnetic spin order, replacing energy-intensive currents with ultra-low-power voltage control. We demonstrate a robust neuromorphic computing device by integrating a collinear antiferromagnet MnIr onto a ferroelectric oxide substrate $0.7PbMg_{1/3}Nb_{2/3}O_3$-$0.3PbTiO_3$ (PMN-PT). By capitalizing on the intrinsic non-volatile memory and strong nonlinearity of this heterostructure, we innovatively introduce an input-modulated *in-situ* self-refreshing encoding mechanism. This paradigm enables direct analog feature extraction without digital preprocessing, thereby circumventing the inherent vulnerabilities of traditional digital front ends. The resulting device exhibits remarkable resilience, withstanding extreme conditions including temperatures up to 500 K, magnetic fields of 55 T, and ionizing radiation doses of 1.5 Mrad (300 rad/s). Leveraging this platform, we achieved 99.8% accuracy in speech recognition, 100% accuracy in multimodal astronaut interaction tasks, and demonstrated autonomous UAV navigation with THz-level response frequencies and an energy consumption of approximately 0.2 fJ per operation, establishing a shielding-free hardware foundation for next-generation deep-space and nuclear applications.

## 2. Results and discussions

### 2.1. Antiferromagnetic neuromorphic platform

The prototype device array features an integrated architecture with word lines (WL) and bit lines (BL) electrodes for applying voltages to the PMN-PT substrate, as shown in the three-dimensional schematic of **Figure 1**a. Through voltage-induced piezoelectric strain, magnetoelastic coupling modulates the antiferromagnetic spin orientation in the MnIr layer, enabling non-volatile information writing and storage via strain-mediated antiferromagnetic-ferroelectric coupling. A photograph of the fabricated device array is presented in Figure 1b. The core modulation mechanism, illustrated in Figure 1c, relies on the reversible tuning of spin alignment in antiferromagnetic MnIr via piezoelectric strain generated by the PMN-PT substrate under an applied electric field, which in turn modulates its electrical resistance (measured via the setup in Figure 1d). Strain-mediated electric-field control of antiferromagnetic order and the associated nonvolatile electrical readout have been established in previous antiferromagnet/piezoelectric heterostructures.[34–39] Building on this physical foundation, the present work focuses on a distinct computing-relevant operating regime, in which the continuously accessible intermediate states, path-dependent resistance evolution, and tunable nonlinear response are exploited as intrinsic computational resources rather than simply as binary or discrete memory states. This electric-field-driven approach circumvents the high current densities required for conventional current-induced torque manipulation, thereby overcoming critical challenges associated with energy consumption and device degradation.

Cross-sectional transmission electron microscopy imaging (Figure 1e) reveals a continuous MnIr layer with a well-defined MnIr/PMN-PT interface, while the MnIr film exhibits a polycrystalline microstructure. The magnetic structure of Mn–Ir alloys is known to depend strongly on composition and crystallographic ordering. Near-equiatomic Mn–Ir compositions are generally associated with the tetragonal $L1_0$-MnIr phase and collinear antiferromagnetic order, whereas the more Mn-rich $L1_2$-$Mn_3Ir$ phase exhibits a coplanar noncollinear triangular antiferromagnetic structure.[40–44] In the present work, the MnIr film was deposited from a nominally equiatomic $Mn_{50}Ir_{50}$ target at 475 °C followed by *in situ* annealing, placing its nominal composition within the reported $L1_0$-MnIr composition range. Although the present structural characterization does not independently quantify the film stoichiometry or degree of $L1_0$ chemical ordering, the magnetic and transport measurements provide additional support for a predominantly collinear antiferromagnetic state. As shown in Figure S1, the measured magnetization contains a linear diamagnetic contribution from the PMN-PT substrate; after subtraction of this background, the MnIr film exhibits nearly vanishing net magnetization (Figure S2). Moreover, the Hall resistance shows a purely ordinary Hall response without a detectable anomalous Hall contribution (Figure S3), in close correspondence with the $L1_0$-MnIr reference characterized in our previous work and distinct from the anomalous Hall and uncompensated magnetic signatures observed in Mn-rich noncollinear Mn–Ir states.[40] The Hall measurements further indicate hole-type carriers (Figure S4a) with a carrier concentration on the order of $10^{23}$ cm$^{-3}$. Taken together, the nominal near-equiatomic composition and the magnetic and Hall characteristics support the assignment of the present MnIr film to a predominantly collinear antiferromagnetic state. Consistent with the strong antiferromagnetic exchange in this material system, $L1_0$-type MnIr has a reported Néel temperature of approximately 1145 K, substantially higher than those of representative antiferromagnets such as $Fe_2O_3$ (950 K) and MnPt (975 K), as summarized in Figure S5.[44–52]

In addition, Figure 1f shows the leakage current of the heterostructure as a function of the applied electric fields, where sharp peaks correspond to polarization switching in the PMN-PT substrate. Notably, the leakage current consistently remains on the order of nanoamperes, effectively suppressing Joule heating during device operation. Throughout the measurements, the negative electric field was maintained at –4 kV/cm, while the positive field was varied from 1.5 to 4 kV/cm. Figure 1g presents the cyclic resistivity-electric field curves under different electric field amplitudes, demonstrating highly reproducible modulation, amplitude-dependent resistance tuning, and stable multi-level electrical responses. These nonlinear dynamics form the physical basis for subsequent self-refreshing encoding and analog computing operations. To clarify the origin of the electric-field-dependent resistance modulation, possible substrate- and measurement-related contributions were also considered. MnIr exhibits metallic conduction with a high carrier concentration, such that electrostatic carrier modulation is strongly screened and is therefore expected to contribute only weakly to the overall resistance change. Likewise, the relatively small strain transferred from PMN-PT is expected to induce only a limited conventional nonmagnetic piezoresistive contribution. In addition, all longitudinal resistance measurements were performed using a four-probe configuration, effectively eliminating contact resistance from the measured signal. In contrast, the ferroelectric/ferroelastic domain switching of PMN-PT is an intrinsic part of the strain-mediated mechanism: it generates distinct remanent strain states that are transferred to MnIr and modify its magnetoelastic anisotropy and antiferromagnetic configuration.[35–40] Therefore, the hysteretic and nonvolatile resistance response in Figure 1g is predominantly attributed to strain-mediated modulation of the MnIr antiferromagnetic state, rather than to interfacial electrostatic effects, contact resistance, or other parasitic contributions. The environmental tolerance of our device is summarized in Figure 1h, which maps its performance across temperature, magnetic field, and radiation dose. The device maintains stable functionality under extreme conditions, including temperatures up to 500 K, magnetic fields of 55 T, and intense radiation exposure, exceeding the operational requirements of harsh environments such as Jupiter-orbit missions, particle accelerators, and fusion reactors. Consequently, the platform is well suited for implementing neuromorphic computing in demanding applications such as voice recognition, pattern recognition, and autonomous systems in extreme space environments (Figure 1i).

**Figure 2** systematically verifies the strain-mediated modulation mechanism, the reversibility of spin-order tuning, and the robustness of the antiferromagnetic device under extreme conditions. We first investigated the electric-field-induced lattice changes in PMN-PT using X-ray diffraction. The (004) diffraction peak was measured successively at applied electric fields of 0, 1.5, 2.5, 3.0, and 4.0 kV/cm. As shown in Figure 2a, a clear and continuous shift in the diffraction peak is observed with increasing electric field. Compared to the initial state, the applied electric field induces an elongation of the out-of-plane lattice constant of PMN-PT, corresponding to an induced in-plane compressive strain. At the maximum field of 4.0 kV/cm, the out-of-plane tensile strain was measured to be 0.073%, from which an in-plane compressive strain of -0.048% is calculated.[53] To further demonstrate the impact of electric-field-induced piezoelectric strain on the antiferromagnetic MnIr, we fabricated a Pt/$Co_{90}Fe_{10}$ (CoFe)/MnIr/PMN-PT heterostructure and characterized its magnetic properties under different applied electric fields (Figure S4b). In contrast to a standalone MnIr thin film (Figure S1), the heterostructure exhibits well-defined magnetic hysteresis loops with a horizontal shift, indicative of exchange bias arising from interfacial coupling between the antiferromagnetic MnIr and the ferromagnetic CoFe layer (Figure 2b). Notably, under an electric field of 4.0 kV/cm, the coercive field ($\mu_0H_C$) and exchange bias field ($\mu_0H_{EB}$) are measured to be 4.9 mT and 8.9 mT, respectively. When an electric field of −1.5 kV/cm is applied, these values change to 5.8 mT and 14.4 mT. Moreover, $\mu_0H_C$ and $\mu_0H_{EB}$ exhibit reversible and repeatable modulation under cycled electric-field switching. Since our previous work has demonstrated that the magnetic properties of the ferromagnetic CoFe layer remain unaffected by piezoelectric strain [35], the observed exchange-bias modulation originates from changes in the antiferromagnetic state of MnIr. Although exchange bias generally originates at the ferromagnet/antiferromagnet interface, the interfacial MnIr spins are exchange-coupled to the antiferromagnetic spin configuration in the interior of the MnIr film. Consequently, the reversible modulation of exchange bias supports strain-induced modification of the antiferromagnetic state of MnIr. Therefore, the reversible modulation of exchange bias provides clear magnetic evidence that the piezoelectric strain modifies the antiferromagnetic configuration of MnIr, which in turn alters the exchange coupling at the MnIr/CoFe interface.

The combined structural, magnetic, and transport results establish a consistent correlation between the applied electric field, the antiferromagnetic order of MnIr, and the resulting resistance modulation. The applied electric field changes the ferroelectric/ferroelastic state of PMN-PT and produces a corresponding lattice strain, which is transferred to the MnIr layer and modifies its magnetoelastic anisotropy. This strain can alter the energetically preferred orientation and/or domain population of the MnIr Néel order. Since the electrical transport of metallic antiferromagnets is sensitive to the Néel-order configuration, such a reorientation or redistribution of the antiferromagnetic order can consequently lead to a change in longitudinal resistivity.[35,40] The hysteretic and nonvolatile resistance modulation observed in Figure 1g can therefore be understood within this strain-mediated pathway, in which the applied electric field is converted successively into piezoelectric strain, antiferromagnetic-order modulation, and an electrical resistance response.

We further investigated the effect of radiation on device performance using $\gamma$-ray irradiation at a dose rate of 300 rad(Si)/s. The resistance states were measured under applied electric fields within one hour following each irradiation step. As shown in Figure 2c, the device exhibits excellent resistivity stability under total ionizing dose (TID) irradiation. Even at a TID of 1.5 Mrad (Si), the relative change in resistivity ($\Delta\rho/\rho_0$) remains below 0.04% for all tested electric fields ($E_G$= −1.5, +1.8, and +4.0 kV/cm), which correspond to the distinct resistive states identified in Figure 1g. These results highlight the strong tolerance of the device to deep-space ionizing radiation. Figure 2d evaluates the device stability under high magnetic fields. After switching by electric fields of −1.5 and +4 kV/cm, the resistivity shows only minor fluctuations as the magnetic field increases up to 55 T. The observed signal jitter is partly attributable to the meter accuracy under ultra-strong magnetic field measurement conditions, confirming the excellent immunity of the device to magnetic interference. In Figure 2e, the thermal stability of the device is evaluated over a broad temperature range up to 500 K. The programmed resistance states remain well preserved throughout the measurement, consistent with the high Néel temperature and robust antiferromagnetic order of MnIr. Notably, the MnIr film exhibits a weak increase in resistivity upon cooling, rather than the conventional positive temperature coefficient expected for a well-ordered bulk metal. Similar behavior has been reported in nanoscale MnIr-based films.[54,55] In the present device, the approximately 10-nm-thick MnIr layer exhibits a polycrystalline microstructure (Figure 1e), for which grain-boundary scattering, structural disorder, and finite-size effects can produce a large residual-resistivity contribution and suppress the conventional electron–phonon-dominated temperature dependence.[56,57] The observed weak negative temperature coefficient can therefore be reasonably understood as a disorder- and finite-size-related metallic thin-film transport behavior, rather than evidence for semiconducting conduction. Furthermore, the long-term reliability is demonstrated in Figure 2f, which shows negligible variation in resistivity over a five-month period following electric-field modulation, underscoring the potential of our device for durable operation in practical settings.

## 2.2. Preprocessing-free all-analog computing in extreme environments

The intrinsic nonlinear dynamics and memory properties of antiferromagnetic computing enable a paradigm shift in physical speech recognition, achieving state-of-the-art performance without any digital preprocessing or encoding, a capability rarely demonstrated in current computing systems that typically rely on extensive front-end signal processing. Here, "preprocessing-free" computing represents a significant departure from conventional physical computing paradigms. While most existing physical learning systems require extensive digital preprocessing—such as time-multiplexing or multiplication by binary random masks—to artificially expand the signal dimensionality before physical processing, our platform entirely bypasses this digital overhead. **Figure 3**a-c employ Uniform Manifold Approximation and Projection (UMAP) dimensionality reduction to visualize the feature extraction capability of physical computing. Prior to processing (Figure 3a), raw speech signals from the TI-46-Word dataset exhibit extreme chaos in UMAP space, with feature points of different digits highly entangled. This behavior indicates not only linear inseparability but also the difficulty of constructing even simple nonlinear decision boundaries. After processing via Input-Modulated *in-situ* Self-Refreshing Encoding (Figure 3c), the raw analog signals are modulated by heterogeneous DC biases that drive the MnIr devices along distinct, history-dependent minor hysteresis loops. As a result, the signals undergo a qualitative transformation in which different digit categories form distinct and tightly clustered groups, with spatial separability emerging purely from device-level processing without any additional computational operations. This demonstrates that MnIr devices successfully map temporally overlapping signals into a high-dimensional, sparse, and linearly separable feature space, thereby achieving computational disentanglement that would otherwise require substantial power consumption in conventional digital signal processors.

The hardware implementation of this encoding mechanism is illustrated in Figure 3b. Here, raw analog speech voltages are directly loaded onto a multi-channel MnIr physical array without digitization. To induce feature diversity, channels are differentiated by applying distinct DC bias voltages $E_{bias}$. Crucially, these biases serve to constrain each device to evolve along a specific, history-dependent minor hysteresis loop (see the detailed processing protocol in Note S1 and Figure S6). Consequently, the output resistance state is governed by the dynamic evolution function $R_{out} = f(R_{current}, E_{in}, E_{base}, E_{bias})$, where $R_{current}$ embodies the magnetic domain history, $E_{in}$ represents the instantaneous analog input voltage, $E_{bias}$ denotes the channel-specific DC offset, and $E_{base}$ refers to the device-specific switching threshold that defines the initial hysteresis state. This architecture effectively operates as a heterogeneous nonlinear filter bank with intrinsic fading memory, in which different bias conditions give rise to distinct nonlinear transformation functions that simultaneously encode instantaneous signal amplitude and temporal context into the analog resistance domain. To validate that this channel diversity stems from device physics rather than mathematical operations, frequency-domain harmonic analysis confirms that variation of the operational hysteresis curve inherently generates distinctly tunable high-order harmonic distortion patterns (Figure S7).

Figure 3d quantitatively evaluates the system performance through a series of systematic comparative experiments, revealing a clear progression from baseline raw signal processing (23.4% accuracy) and a pure digital convolutional neural network (CNN, 73.2%) to physically enhanced computing with increasing channel counts. We compare raw and device-processed signals using identical backend capacities to separate the device's contribution from the algorithm. With a pure linear classifier, a single physical channel more than doubles the baseline accuracy to 54.6%. Furthermore, using a single-layer CNN, both the raw and single-channel inputs utilize exactly 524 training parameters; yet the physical preprocessing leads to a 19% improvement in accuracy, despite a 5-fold temporal downsampling. Specifically, a single physical channel achieves an accuracy of 92.8%, which increases to 95.0% with 8 channels, 97.0% with 16 channels, and 97.2% with 32 channels. The ultimate configuration, which combines 32-channel physical feature extraction with a two-layer CNN backend, achieves a near-perfect recognition accuracy of 99.8% (detailed algorithmic implementation in Note S2 and Figure S8). Notably, all convolutions and full connections are also implemented using the non-volatile resistive states of MnIr devices (specific implementation methods can be found in Note S3 and Figures S8-S11). The corresponding training convergence curves, comparing the full-precision baseline with the hardware implementation, as well as the detailed confusion matrix from ten-fold cross-validation, are illustrated in Figures S12-S13. The radar chart in Figure 3e summarizes the comprehensive advantages of the proposed system across multiple critical metrics. In particular, with respect to the architectural elimination of digital preprocessing, this work demonstrates a capability that remains largely inaccessible to existing physical computing platforms, thereby establishing a foundation for truly *in-situ* analog computing. More critically, the antiferromagnetic system exhibits consistently strong performance across environmental resilience and computational accuracy, including an operating temperature range of 1.6–500 K, magnetic-field tolerance up to 55 T, radiation hardness of 1.5 Mrad, and a recognition accuracy of 99.8%. Together, these attributes establish a new benchmark for all-capable computing solutions in extreme environments that comprehensively surpasses silicon-based MEMS, oxide memristors, spintronic oscillators, and conventional analog circuits (see detailed comparison in Table 1).[58–62]

Compared with other physical computing platforms, the present device offers strong and electrically tunable nonlinearity together with nonvolatile channel configurability. Its operation is not constrained by fixed physical relaxation time constants, enabling the processing of analog signals across diverse timescales. The device memory instead originates from nonvolatile domain path dependence rather than leaky temporal integration. Consequently, it does not inherently provide the exponentially fading short-term memory characteristic of dedicated reservoir systems, and an external refresh operation is introduced for strongly history-dependent temporal tasks (Figures S14-S15). A detailed comparison along four axes is provided in Table S1.

Beyond speech recognition, we further demonstrate the universality of this physical computing paradigm through chaotic time series prediction using the Mackey-Glass benchmark (Figure S14). In this implementation, eight MnIr devices are configured as a physical dynamic processor that directly processes the input sequence through random voltage projection. The intrinsic hysteretic dynamics and path-dependent state evolution of the devices naturally provide the nonlinear temporal kernel required for modeling chaotic dynamics. After normalization, the resistance states are linearly read out using a linear layer, enabling accurate multi-step prediction with a mean squared error on the order of $10^{-3}$ level. These results validate that antiferromagnetic/ferroelectric heterostructures can function as universal nonlinear dynamic processors for diverse time-dependent computational tasks without task-specific architectural reconfiguration. Furthermore, we demonstrate spacecraft angular velocity denoising using corrupted measurements from the Juno probe, fused with magnetic field data as six-dimensional inputs to a parallel MnIr processing array. (Figure S15 and Note S4). Random projection matrices drive each device channel along distinct voltage-dependent hysteresis trajectories, generating high-dimensional nonlinear features that are subsequently concatenated with the raw inputs for state augmentation. These augmented states are then linearly decoded via ridge regression to reconstruct the clean angular velocities. This result validates the capability of the antiferromagnetic system for robust multi-sensor fusion and signal processing in extreme space environments, where conventional digital preprocessing would be highly vulnerable to radiation-induced failure.

**Figure 4** illustrates the capability of the all-analog processor to perform complex, multi-task *in situ* processing under simulated catastrophic space-station conditions, including intense radiation and magnetic-field interference associated with severe solar storms. In scenarios where conventional digital computing systems would be compromised by radiation-induced single-event upsets and logic failures, the antiferromagnetic MnIr processor autonomously executes three critical tasks in parallel: visual object recognition (Object), gesture command recognition (Gesture), and real-time action capture (Action) (Figure 4a). Crucially, this architecture establishes a truly preprocessing-free, all-analog pipeline. Raw sensor signals directly modulate the strain-mediated magnetic states, allowing the device's intrinsic nonlinear dynamics to physically execute feature extraction and classification. Such multi-modal processing capability provides a technological foundation for autonomous decision-making in extreme environments where traditional silicon-based systems are unable to operate reliably. Detailed network architectures and implementation methodologies for these multimodal tasks are comprehensively described in Note S5.

Specifically for the action capture task, real-time encoding of dynamic action sequences is achieved through distributed nonlinear computing across an array of eighteen antiferromagnetic MnIr devices (Figure 4b). Eighteen-dimensional skeletal keypoint data extracted from human actions are directly applied to the device array as analog voltage signals (representative skeletal sequences are visualized in Figure S16), where the intrinsic nonlinear computing dynamics naturally capture temporal features. During the 125-frame temporal input process, the non-volatile resistance states of the devices continuously evolve over time, as shown by the heatmap in Figure 4b, converting complex spatiotemporal action features into high-dimensional resistance trajectories. The confusion matrix in Figure 4c indicates that the system reliably distinguishes eight critical commands, including wave, jogging, and circle, with an overall accuracy of 95.3% and near-unity recognition accuracy for several action categories, demonstrating the powerful capability of the physical computing for capturing temporal dynamic features.

Figure 4d emphasizes the inherent advantages of the proposed system in processing high-noise signals encountered in extreme radiation environments. The nonlinear saturation characteristics of MnIr devices, arising from the magnetoelastic coupling properties, naturally serve as analog denoising and feature enhancement mechanisms without the need for additional digital signal processing stages. Blurred object images contaminated with noise (examples of low-contrast and noisy inputs are shown in Figure S17) undergo physical layer processing, where nonlinear transfer functions effectively suppress high-frequency noise components while significantly enhancing key object features through analog domain filtering. This physical denoising mechanism enables the backend classifier to maintain 100% classification accuracy in a 10-class visual object recognition task (Figure 4e), demonstrating robustness that surpasses traditional digital filtering approaches under extremely low signal-to-noise ratio environments typical of radiation-compromised imaging systems. Figure 4f, g further demonstrates the capability of the system to directly process analog signals from inertial sensors. Specifically, a wearable MPU-6050 accelerometer attached to a human hand collects real-time three-axis acceleration data. The dynamic motion signals are directly mapped into a high-dimensional feature space and input as voltage signals to the MnIr device array. The MnIr processor autonomously distinguishes five distinct postures—up, down, left, right, and stationary states—with an average accuracy of 98%, thereby realizing *in situ* fusion and decision-making based on multi-source sensory data. The training dynamics for these multimodal tasks exhibit rapid convergence and stable learning behavior, as illustrated by the accuracy and loss curves in Figure 4h, i.

To further establish the universality of this computing paradigm beyond specific sensor modalities, we benchmarked the system on standard computer vision tasks, including MNIST and Fashion-MNIST. In this architecture, the input data first undergoes nonlinear feature extraction via the MnIr devices, and is subsequently processed by a linear classification layer composed of the MnIr crossbar array using differential conductance weights. This all-analog pipeline achieved robust recognition accuracies of 91.4% for handwritten digits (Figure S18) and 80.63% for fashion articles (Figure S19). Notably, these hardware-implemented results exhibit negligible performance degradation compared to full-precision software baselines, validating the fidelity and effectiveness of the analog computing architecture.

The respective contributions of the material platform and the computing strategy can be clearly distinguished. At the material level, the MnIr/PMN-PT heterostructure provides electrically tunable nonlinearity, nonvolatile and continuously distributed resistance states along amplitude-dependent minor loops, and nA-level leakage currents that enable continuous analog waveforms to be directly applied with minimal Joule heating. It also offers robust operation under extreme conditions, including high temperature, ultra-high magnetic fields, and ionizing radiation. At the computing level, the proposed architecture exploits these intrinsic device characteristics through a heterogeneous multi-channel configuration strategy, direct input mapping without time-frequency transformation, and a non-uniform quantization-aware training scheme that maps backend weights onto experimentally accessible conductance states. The backend readout follows the established protocol commonly used in physical computing systems.

### 2.3. All-hardware integrated autonomous drone vision cruise system

**Figure 5** presents system-level validation in which the MnIr neuromorphic chip is integrated as the core central processing unit onto a drone printed circuit board (Figure 5a), representing a decisive step towards replacing traditional digital flight controllers with analog processors for closed-loop autonomous navigation and control under GPS-denied conditions. The complete hardware signal flow, illustrating the autonomous interaction between the onboard control modules and the integrated antiferromagnetic processing unit to achieve closed-loop operation without external computing support, is depicted in Figure S20. Figure 5b illustrates the complete visual navigation processing pipeline. Ground-marker images captured by the onboard camera are resized to a resolution of 100×100 pixels and partitioned into 10×10 pixel patches using a stride of (10, 10), generating 100 sequential patch inputs. These patches undergo self-refreshing feature mapping by being input as temporal voltage signals to the MnIr device array, where pixel intensity directly drives nonlinear evolution of device resistance states. This patch-based processing employs a parallel spatiotemporal coding strategy that preserves local spatial correlations while leveraging intrinsic temporal dynamics of the devices for efficient feature mapping (Note S6). A four-step feature fusion strategy merges responses from every four consecutive patches into one feature vector, generating high-dimensional resistance feature maps that are flattened and input to a linear classification layer, directly outputting seven flight control commands including forward, backward, left turn, right turn, descend-and-spin, and hover-and-land, with the entire pipeline requiring no traditional CPU intervention. A summary of the datasets, hardware configurations, computational tasks, and evaluation metrics of all demonstrated applications is provided in Table S2.

Figure 5c demonstrates system robustness through training on Easy (forward-placed markers) and Hard (random rotation angles 0-40°) datasets (examples in Figure S21). The Easy dataset rapidly converges within 20 training epochs with accuracy approaching 100%. In contrast, the Hard dataset, despite initial fluctuations, ultimately converges to extremely high recognition accuracy reaching 82%, demonstrating effective extraction of rotation-invariant features through the nonlinear feature mapping capability of MnIr devices. The actual flight photo sequence in Figure 5d (complete process in Supplementary Video 1) displays complex maneuvers under fully autonomous control. During the cruise phase, the system executes left-turn, right-turn and backward commands based on arrow markers, while during task triggering, the second triangular markers initiate descend and spin commands for target inspection. After successfully capturing all four types of patterns twice, the drone automatically returns to the H-shaped area and completes landing. Actual measurements indicate real-time marker identification with latency under 50 milliseconds; in this system, the MnIr array performs the nonlinear feature mapping, while the host computer performs image acquisition, patch generation, and readout (Figure S20). Finally, Figure 5e summarizes the overall system performance using a three-dimensional scatter plot. Compared with previously reported neuromorphic solutions,[60,63–70] the present system exhibits a clear advantage in environmental resilience, including operating temperature range, magnetic-field tolerance, and radiation hardness, while simultaneously maintaining an ultrahigh device response frequency exceeding ~1 THz and an ultra-low single-operation energy consumption of ~0.2 fJ per operation (see Note S7 for calculation details and see detailed comparison in Table 2). Detailed performance comparisons regarding energy consumption and response frequency with other state-of-the-art neuromorphic devices are listed in Table 2. Collectively, these results establish the feasibility of replacing traditional digital controllers with antiferromagnetic analog processors in extreme-environment missions, such as deep-space exploration, nuclear facility monitoring, and high-energy physics experiments, enabling reliable *in situ* autonomous computing beyond the limits of conventional electronics.

**3. Conclusion**

In summary, this work demonstrates that antiferromagnetic analog computing provides a viable solution to the fragility bottleneck that constrains edge intelligence in extreme environments. The MnIr/PMN-PT neuromorphic processor simultaneously tolerates temperature up to 500 K, magnetic fields of 55 T, and radiation doses of 1.5 Mrad, enabled by an input-modulated in-situ self-refreshing encoding scheme that eliminates the need for digital preprocessing. By exploiting the nonlinear, history-dependent resistance dynamics of antiferromagnetic domains for direct analog feature extraction, the system achieves a speech recognition accuracy of 99.8% and delivers high-fidelity multi-modal recognition for mission-critical astronaut monitoring, and supports real-time autonomous drone navigation with an energy consumption of approximately 0.2 fJ per operation, representing orders of magnitude improvements in efficiency over conventional digital signal processors. Taken together, the combination of exceptional environmental resilience, a preprocessing-free architecture, and competitive computational performance establishes antiferromagnetic analog computing as a promising pathway for mission-critical applications, including deep-space exploration, fusion energy monitoring, and high-energy physics instrumentation, where conventional silicon electronics are unable to operate reliably.

**4. Experimental Section**

*Materials Growth*

MnIr thin films were deposited on (001)-oriented $0.7PbMg_{1/3}Nb_{2/3}O_3-0.3PbTiO_3$ (PMN-PT) substrates by a *d.c.* magnetron sputtering system utilizing a $Mn_{50}Ir_{50}$ polycrystalline target. The sputtering system was evacuated to a base pressure of $7.5 \times 10^{-9}$ Torr prior to deposition. Film growth was carried out at 475 °C with a sputtering power of 60 W under an argon pressure of 3 mTorr. The distance between the target and the substrate was held constant at 245 mm. After deposition, the samples were *in situ* annealed for 30 mins and then gradually cooled down to room temperature. For the $Pt/Co_{90}Fe_{10}$/MnIr/PMN-PT heterostructures, the $Co_{90}Fe_{10}$ and Pt layers were deposited after the growth of MnIr layer at room temperature by *d.c.* sputtering under an argon pressure of 3 mTorr. The $Co_{90}Fe_{10}$ layer was grown at 90 W with a rate of 0.29 $Å \cdot s^{-1}$, while the Pt layer was deposited at 30 W with a rate of 0.28 $Å \cdot s^{-1}$.

*Device Fabrication*

The device array was fabricated from MnIr/PMN-PT heterostructures. MnIr arrays were patterned using maskless ultraviolet lithography and ion beam etching. An insulating MgO layer was employed to prevent electrical short circuits. The top and bottom electrodes, consisting of Cr (10 nm) and Au (45 nm), were deposited via electron beam evaporation.

*Structure Characterization*

X-ray diffraction measurements were conducted using a Bruker D8 Advance diffractometer with Cu-$K_\alpha$ radiation ($\lambda$ = 1.5418 Å) operating at 40 kV and 40 mA. Cross-sectional transmission electron microscopy (TEM) samples were prepared from the MnIr/PMN-PT heterostructures by focused $Ga^+$ ion beam milling. The subsequent TEM measurements were carried out on a JEOL NeoArm system at an operating voltage of 200 kV.

*Electrical and Magnetic Measurements*

The longitudinal resistance was measured using the four-probe method within a Quantum Design physical property measurement system (PPMS). Electrical contacts were made to the electrodes via an Al wire bonder. A Keithley 6221 source meter supplied the measurement current, while the corresponding voltage was detected with a Keithley 2182A nanovoltmeter. An external electric field was applied using a Keithley 2400 source meter. The magnetic properties were characterized using a Quantum Design VersaLab system, with magnetization data acquired via its vibrating sample magnetometer (VSM) module at various temperatures and magnetic fields.

*High magnetic field transport measurements*

The measurements were performed at the Wuhan National High Magnetic Field Center. Electrical contacts in a linear four-probe configuration were established with silver paint and Au wires. The longitudinal resistance was measured using a linear four-probe method. A 500 μA *a.c.* current (89 kHz) from a National Instruments PXI-5402 generator was applied, and the voltage was recorded by a National Instruments PXIe-5105 oscilloscope (4 MHz sampling), with the signal pre-amplified 100 times using a Stanford Research SR560 preamplifier for enhanced accuracy.

*γ-Ray irradiation on devices*

Irradiation was performed using a Cobalt-60 ($^{60}$Co) $\gamma$-ray source. The radioisotope has a half-life of 5.27 years and decays to stable $^{60}$Ni, producing $\gamma$-ray photons in the process ($^{60}_{27}Co \rightarrow {}^{60}_{28}Ni + {}^{0}_{-1}\beta + {}^{0}_{0}\gamma$). A fixed dose rate of 300 rad(Si)/s was applied, ensuring that the total ionizing dose (TID) scaled linearly with exposure time. Electrical characterization of the devices, specifically the voltage-dependent longitudinal resistance, was conducted within one hour after each irradiation step.

*Input-Modulated In-situ Self-Refreshing Encoding*

The core self-refreshing mechanism was implemented by exploiting the hysteretic resistance response of the MnIr devices. The device's resistance state is not a static function of input voltage but depends on the historical trajectory on the hysteresis loop. When an analog voltage signal is applied, it dynamically drives the operating point across different minor loops depending on whether the input exceeds specific threshold voltages. To realize multi-channel feature extraction, we configured 16-32 parallel physical channels by applying distinct DC bias voltages and scaling factors to each device, thereby creating a heterogeneous array of non-linear filters that naturally separate entangled temporal features in the analog domain.

*Signal processing and network implementation*

For speech recognition (TI-46-Word), raw analog signals were directly fed into the 32-channel MnIr array. The physically extracted features were then processed by a hardware-based backend comprising a single-layer 1D CNN (kernel size 3, 32 output channels) and a linear classifier, implemented using the same MnIr devices as synaptic weights. For action recognition, 18-dimensional skeletal coordinates were mapped to 18 physical nodes. The temporal evolution of resistance states over 125 frames were aggregated to form a spatiotemporal feature vector, which was classified by the Hardware-constrained MnIr synaptic layer.

*Drone vision system integration*

To realize all-hardware autonomous control, we physically integrated a custom neuromorphic PCB hosting MnIr device onto a drone platform, serving as the core processing unit to replace traditional digital controllers. During flight, the onboard camera captures ground marker images at 30 fps. To efficiently process high-dimensional visual data, we employed a time-multiplexing strategy where the image is segmented into sequences of 10×10 pixel patches. These patches are converted into analog voltage signals and sequentially input into the reusable MnIr device array. Leveraging the "Input-Modulated In-situ Self-Refreshing Encoding" mechanism, the pixel intensity of each patch directly drives the nonlinear evolution of the device resistance states, thereby completing feature mapping in the analog domain. The recognition results directly generate flight commands (e.g., forward, turn, hover), which are fed back to the drone's flight controller, achieving closed-loop autonomous visual cruising.

**Data Availability Statement**

The data that support the findings of this study are available from the corresponding authors upon reasonable request.

**Supporting Information**

Additional supporting information can be found online in the Supporting Information section.

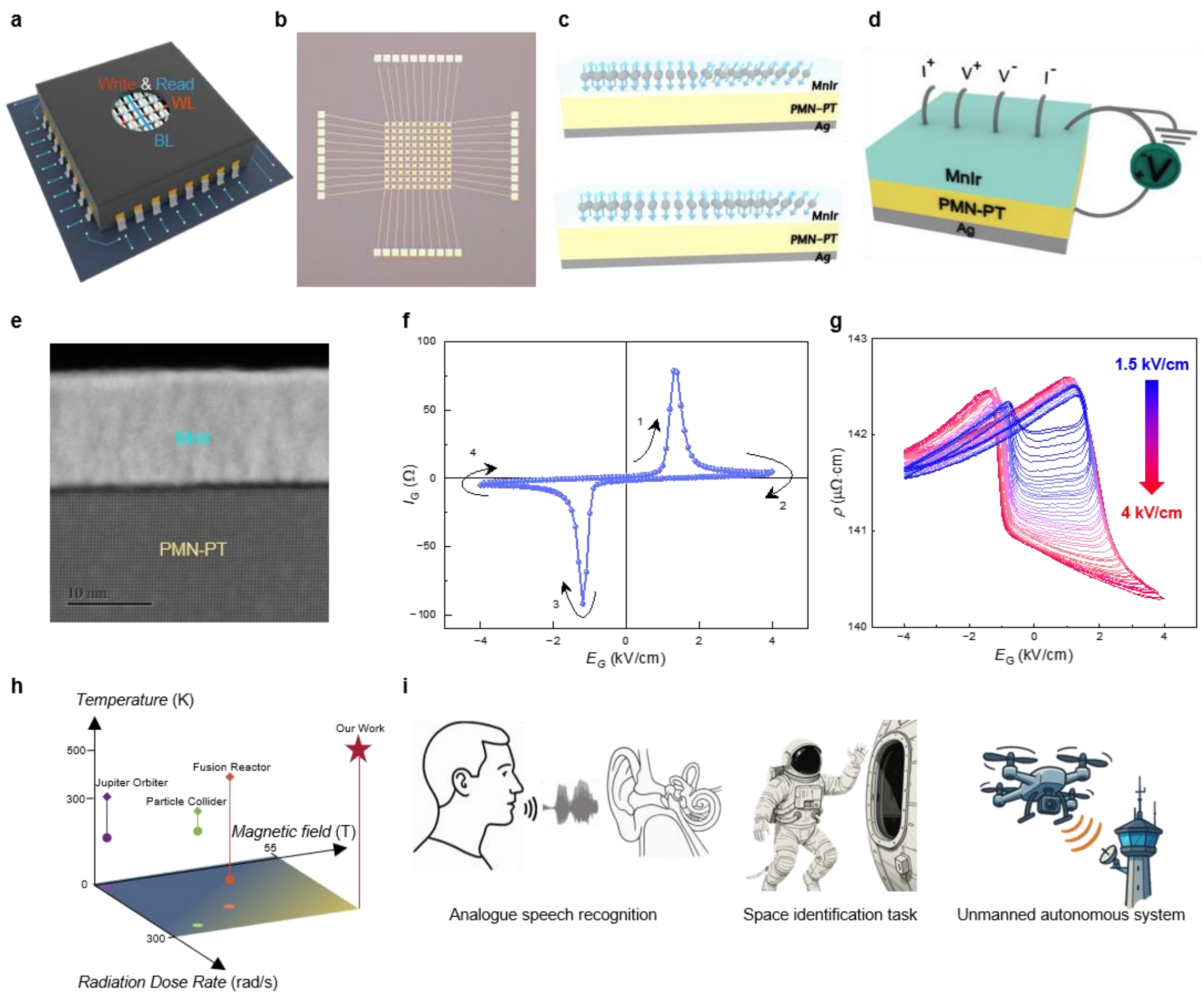

**Figure 1 | Physical foundation and extreme environment benchmarks of the antiferromagnetic neuromorphic processor.** (a) Three-dimensional schematic of the MnIr/PMN-PT device array architecture, showing integrated WL and BL for applying gate electric fields. (b) Optical microscopy image of the MnIr/PMN-PT device array. (c) Illustration of the strain-mediated switching mechanism where piezoelectric strain from the PMN-PT substrate modulates the antiferromagnetic spin orientation of the MnIr layer. (d) Schematic of the geometry for measuring the resistance of the MnIr/0.7$PbMg_{1/3}Nb_{2/3}O_3$-0.3$PbTiO_3$ (PMN-PT) heterostructure under electric fields. (e) Cross-sectional HRTEM image of the MnIr/PMN-PT interface. (f) Current-electric field characteristics showing low leakage current. (g) Resistivity-electric field hysteresis loops measured at room temperature, demonstrating the nonlinear memory switching behavior. (h) Three-dimensional performance benchmark comparing the environmental tolerance of this work with requirements for nuclear fusion reactors, Jupiter exploration, and particle accelerators; Our Work (red star) simultaneously exceeds all thresholds. (i) Application scenarios including deep-space speech recognition, astronaut multimodal sensing, and autonomous drone navigation.

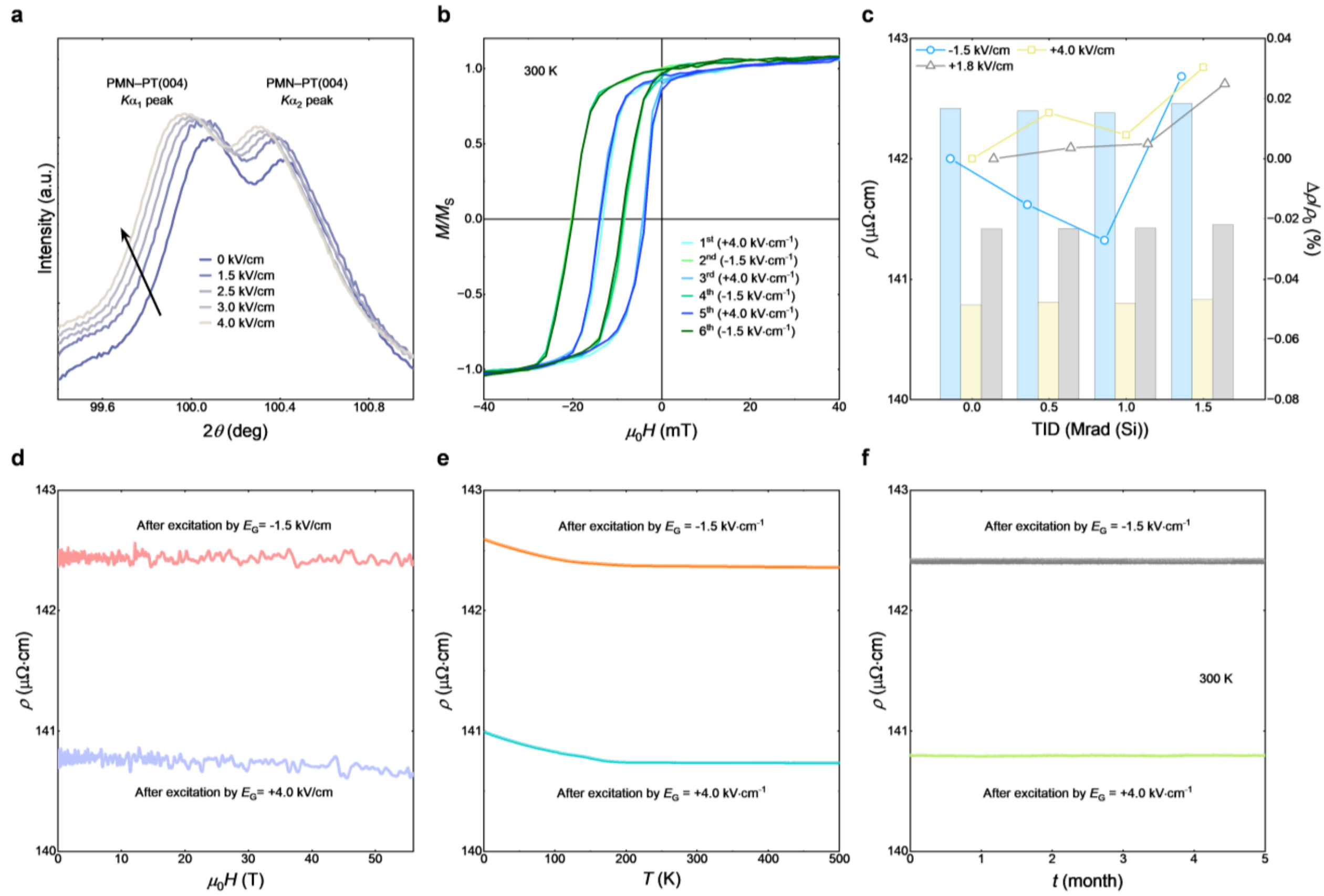

**Figure 2 | Comprehensive device characterization and reliability metrics.** (a) X-ray diffraction patterns of the PMN-PT substrate under different applied electric fields, showing the shift of the (004) peak indicative of lattice elongation. (b) Magnetic hysteresis loops of a reference Pt/CoFe/MnIr/PMN-PT heterostructure under different electric fields, confirming the modulation of antiferromagnetic order via exchange bias shifts. (c) Radiation hardness evaluation showing stable resistivity under gamma irradiation up to 1.5 Mrad (Si equivalent). (d) Magnetic field immunity testing demonstrating preserved resistance states under ultra-high magnetic fields up to 55 T. (e) Temperature-dependent resistivity measurements showing stability from 1.6 to 500 K. (f) Long-term reliability test proving robust data retention exceeding 5 months under ambient conditions.

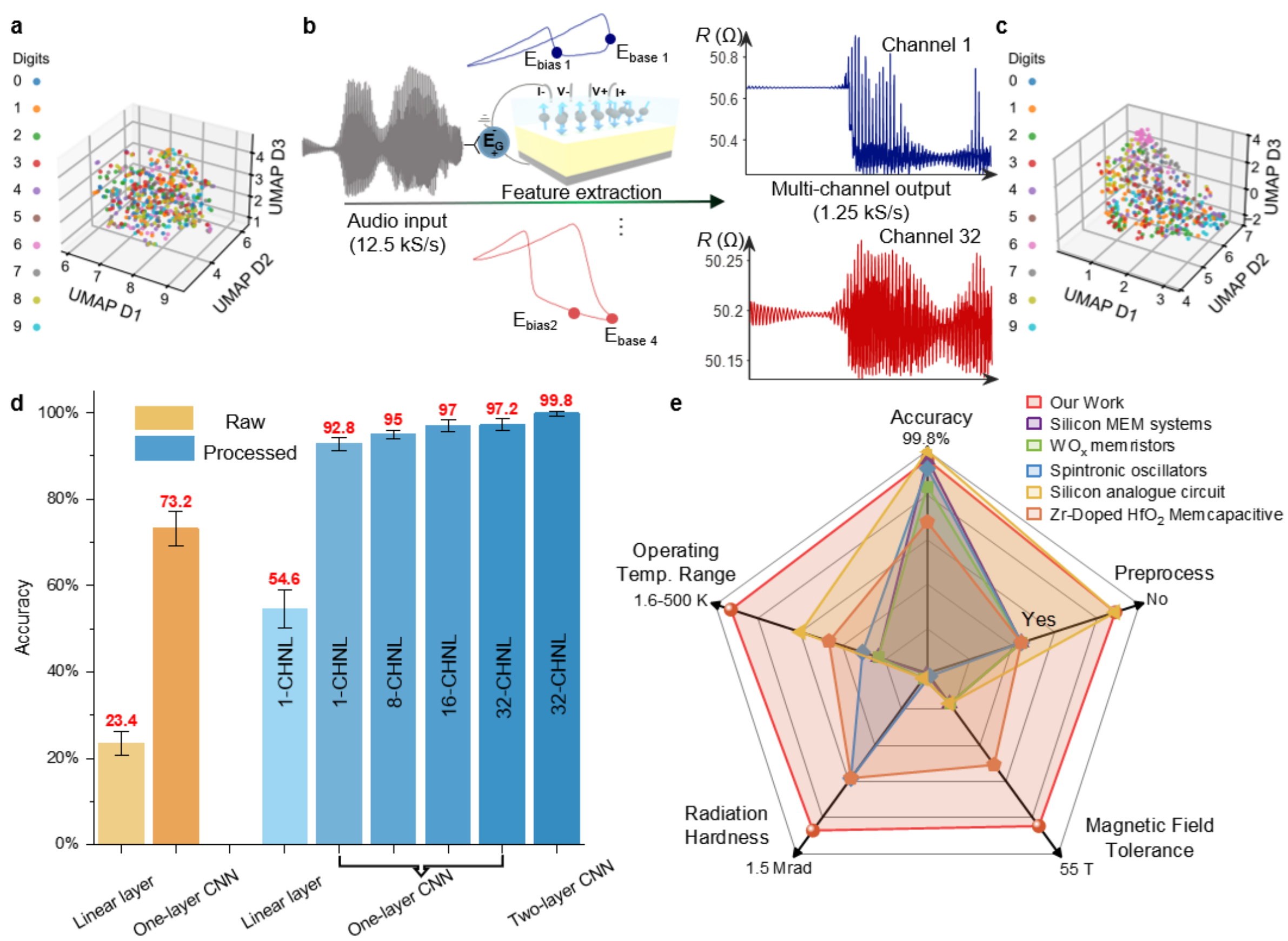

**Figure 3 | ADC-free all-analog speech recognition via Input-Modulated In-situ Self-Refreshing Encoding.** (a) UMAP visualization of raw speech signals from the TI-46-Word dataset, showing highly entangled features before processing. (b) Schematic of the multi-channel analog feature extraction pipeline where different DC bias voltages configure devices as heterogeneous nonlinear filters. (c) UMAP visualization after MnIr device processing, revealing the emergence of distinct, linearly separable clusters for each digit class. (d) Recognition accuracy comparison showing the systematic improvement from a linear baseline (23.4%) to the physics-enhanced performance with increasing channel counts (up to 97.0% with 32 channels), and finally achieving 99.8% with the all-hardware network. (e) Radar chart benchmarking this work against existing technologies across five key metrics: preprocessing elimination, environmental resilience, and accuracy, showing comprehensive superiority (red region).

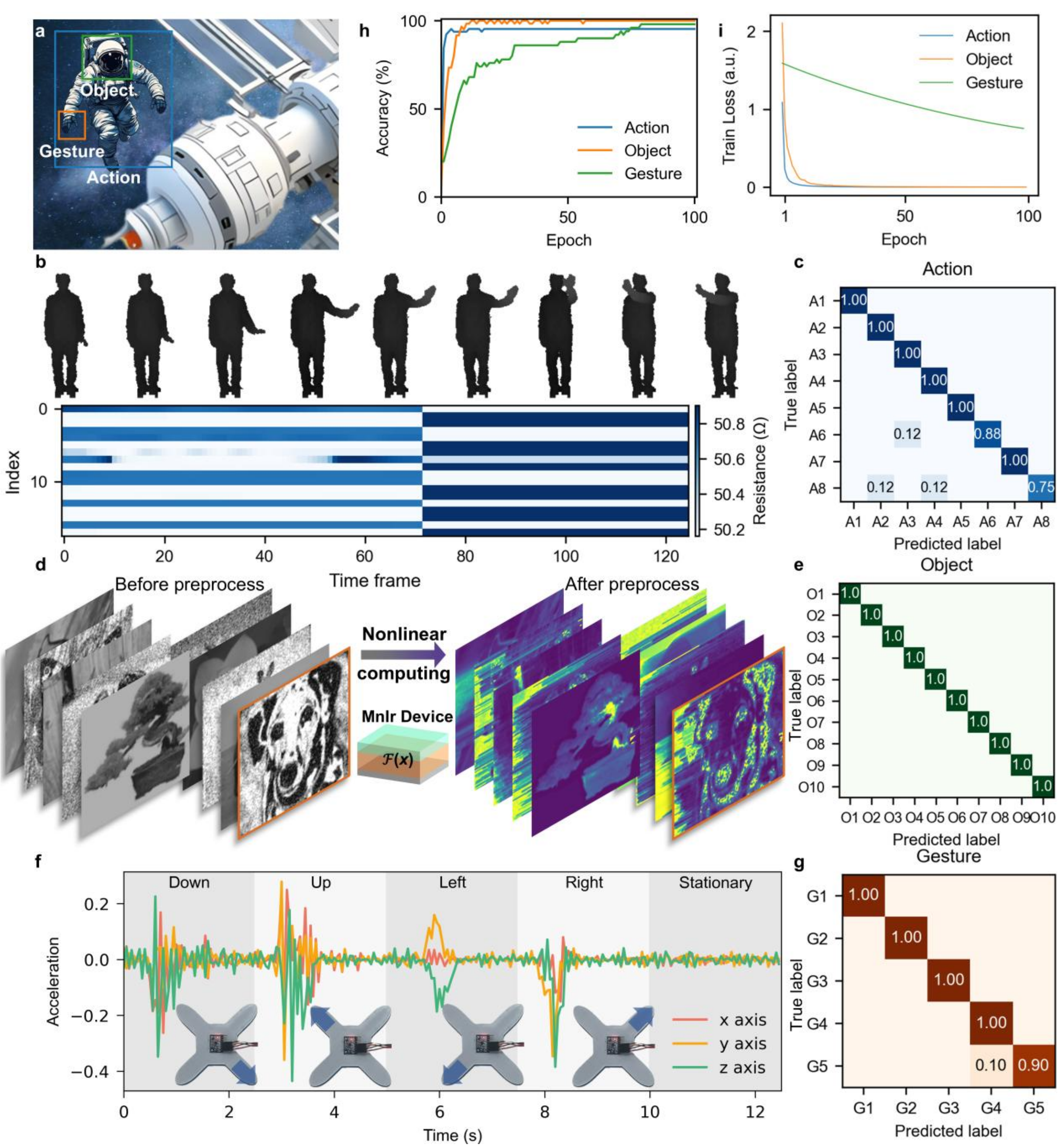

**Figure 4 | Multimodal perception and robust control for extreme space station scenarios.** (a) Schematic of a crisis scenario where the MnIr processor assumes responsibility for onboard visual object, gesture, and action recognition during digital system failure. (b) Visualization of the physical dynamic computing process for action recognition, where 18-dimensional skeletal data is encoded into resistance trajectories by an 18-node physical device. (c) Confusion matrix for the 8-class action recognition task demonstrating nearly 100% accuracy for critical commands. (d) Demonstration of the physical denoising mechanism where the device's nonlinear saturation naturally enhances key object features from radiation-corrupted noisy images (raw images adapted from the Caltech-101 database). (e) Confusion matrix for 10-class visual object recognition achieving 100% accuracy under low signal-to-noise conditions. (f) Real-time processing of analog signals from a 3-axis accelerometer. (g) Confusion matrix for five-class posture recognition (up, down, left, right, stationary) showing precise discrimination.

(h) Test accuracy curves for the three tasks (Action, Object, Gesture). (i) Training loss curves showing convergence behavior.

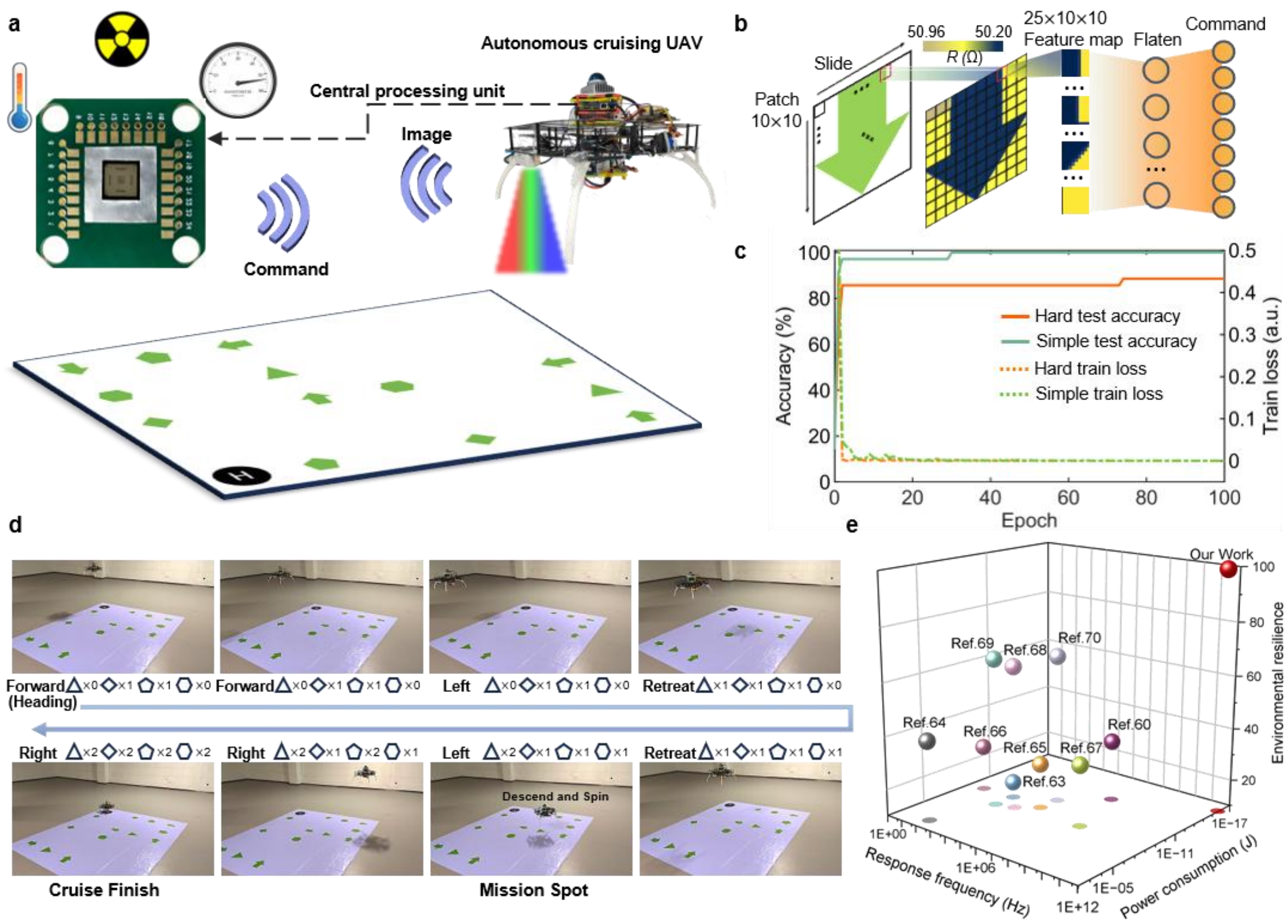

**Figure 5 | All-hardware integrated autonomous drone vision cruise system.** (a) Photograph of the custom PCB hosting the MnIr neuromorphic chip integrated into a drone for GPS-denied autonomous navigation. (b) Analog visual processing pipeline showing patch-based physical convolution and self-refreshing feature mapping to directly convert filmed images into flight commands. (c) Training convergence curves for Easy and Hard datasets, demonstrating robust feature learning. (d) Flight photo sequence showing the successful execution of complex maneuvers including arrow-guided turns and target-triggered landings. (e) Three-dimensional performance comparison plotting environmental resilience versus response frequency and energy efficiency; Our Work simultaneously achieves >THz operation frequency, 0.2 fJ per operation energy consumption, and comprehensive extreme environment tolerance.

**Table 1 | Comparison of recognition accuracy and environmental resilience with state-of-the-art neuromorphic computing systems.**

| System | Test Accuracy (%) | Pre-process Method | Environmental Resilience | | |
|---|---|---|---|---|---|
| | | | Magnetic Field Tolerance (T) | Radiation Hardness (Mrad) | Operating Temp. Range (K) |
| Our work | 99.8 | N/A | >55 | >1.5 | 1.6 - 500 |
| Silicon MEM systems [58] | 99.8 | Cochlear | >10 | >0.01 | 233 - 358 |
| $WO_x$ memristors [59] | 99.2 | Cochlear | >10 | >0.05 | 233 - 358 |
| Spintronic oscillators [60] | 80/99.6 | Spectrogram/Cochlear | Sensitive | >1 | 233 - 398 |
| Silicon analogue circuit [61] | 99.9 | N/A | >10 | >0.05 | 77 - 398 |
| Zr-Doped $HfO_2$ Memcapacitive [62] | 98.4 | Cochlear | >30 | >1 | 200 - 450 |

**Table 2 | Comparison of response frequency, energy efficiency and comprehensive extreme environment tolerance with state-of-the-art neuromorphic computing systems.**

| System | Power Consumption (nJ) | Response Frequency (Hz) | Environmental Resilience | | | |
|---|---|---|---|---|---|---|
| | | | Magnetic Field Tolerance (T) | Radiation Hardness (Mrad) | Operating Temp. Range (K) | Score |
| Our work | $2\times10^{-7}$ | $10^{12}$ (*Adaptable limit*) | >55 | >1.5 | 1.6 - 500 | 100 |
| Organic electrochemical transistors [63] | 0.56 | $10^{3}$ | >10 | >0.1 | 250 – 370 | 16 |
| Self-organizing nanowire networks [64] | $7.5\times10^{5}$ | $10^{2}$ (*~10 ms pulses*) | >20 | >0.5 | 200 – 450 | 40 |
| Magnetic domains [65] | 0.93 | $10^{5}$ | >0.1 | >0.05 | 4 - 398 | 28 |
| Spintronic oscillators [60] | $10^{-4}$ (*Scaled*) | $10^{7}$ (*~100 ns pulses*) | Sensitive | >1.0 | 233 – 398 | 33 |
| Brownian skyrmion reservoir [66] | 0.7 | 10 (*~60 ms frames*) | Sensitive | >1.0 | 285 – 360 | 27 |
| Magnon-scattering reservoir [67] | 5 | $10^{7}$ (*~20 ns pulses*) | Sensitive | >1.0 | 233 – 398 | 33 |
| Dynamic oxide memristor [68] | 10 | $10^{3}$ (*0.2 ms pulses*) | >55 | >1.0 | 250 - 370 | 64 |
| Ferroelectric diode [69] | 23.6 | $10^{2}$ (*~2 ms pulses*) | >55 | >1.0 | 250 – 400 | 66 |
| Ferroelectric MPB transistor [70] | 0.0225 | $10^{4}$ (*~0.1 ms pulses*) | >55 | >1.0 | 250 - 400 | 66 |

Supporting Information

# Bare-Die Antiferromagnetic Computing

*Yu Liu[#], Zhuoting Han[#], Zexin Feng[#], Peixin Qin*, Zhiyuan Duan, Yuhao Ye, Zengwei Zhu, Chengyan Zhong, Li Liu, Guojian Zhao, Wenbin Shen, Jingyu Li, Sixu Jiang, Xiaoyang Tan, Xiaoning Wang, Ziang Meng*, Chengbao Jiang*, Zhiqi Liu**

#These authors contributed equally to this work.

# Notes

### Note S1 Antiferromagnetic nonlinear neuromorphic computing Strategy

Although the antiferromagnetic device possesses non-volatile memory characteristics, the hysteresis effect induced by piezoelectric strain enables it to transform temporal information—encoded as different piezoelectric voltages—into spatial information. As shown in Figure S6, five groups of varying piezoelectric voltages were applied to the device. In each group, only the voltage applied in the 4th round differed. From initialization at the 0th round until just before the 4th round, the MnIr layer in every group exhibited identical resistivity. Upon applying different positive piezoelectric voltages in the 4th round, the ferroelectric domains underwent varying degrees of positive polarization switching, causing variations in the resistivity of the MnIr layers across the groups. In subsequent rounds, the piezoelectric voltage applied to each group was lower than the maximum positive voltage applied in the 5th round. Consequently, the resistivity of the MnIr layer depended not only on the currently applied voltage but also on the past switching states of the ferroelectric domains, thereby exhibiting synaptic memory behavior. This memory effect decays as the subsequently applied piezoelectric voltage decreases. When the voltage is reduced to the negative coercive field of the ferroelectric domains, the switching state of each group returns to the initial state, and the resistivity of the MnIr layers becomes consistent again.

The dynamic process illustrated in Figure S6 underpins the concepts of "fading memory" and "self-refreshing encoding" central to our computing framework. Specifically, "fading memory" naturally arises from the history-dependent paths along the piezoelectric minor hysteresis loops. The current resistance state essentially "remembers" the sequence of recent voltage inputs (e.g., the distinct resistivity variations after the 4th round). However, as new dynamic inputs arrive, they drive the ferroelectric domains into new local minor loops, causing the influence of older states to gradually decay or "fade", which provides the crucial short-term temporal context required for time-series processing. Furthermore, "self-refreshing encoding" refers to the autonomous nature of this state update. In typical memristive systems, maintaining dynamic memory often requires explicitly designed reset pulses to clear old states. In our framework, the continuously varying analog input itself acts as the driving force. The natural voltage fluctuations of the input signal continuously push the device across different minor loops, intrinsically updating (refreshing) the spatial resistance state *in situ* without external digital intervention.

The fading memory exhibited by the device depends on the hysteresis path of the ferroelectric domains. We can reconfigure the device by applying a negative polarization voltage, thereby altering its memory strength to match different temporal tasks. Figure S14 and Figure S15 compare the impact of performing or omitting this refresh operation on the gyroscope angular velocity post-processing task in a Jovian environment and the Mackey-Glass (MG) sequence prediction task. For image processing tasks and analogue speech recognition tasks, performance depends more heavily on the nonlinearity strength of the device. Stronger nonlinearity enables the device to map original image information into a high-dimensional space, thereby extracting more effective image features. Figure 1f demonstrates that the device exhibits strong nonlinearity under any degree of positive polarization. Therefore, we do not apply additional piezoelectric voltage for refreshing in the image processing and analogue recognition tasks.

### Note S2: Network architecture details and training procedures

#### 2.1 Dataset preparation

The speech recognition benchmarks utilized a subset of the TI-46-Word Speech Dataset. We selected recordings from 5 speakers for the 10-digit subset ("zero" through "nine"), with each speaker providing 10 utterances per digit, yielding 50 samples per class and a total dataset size of 500 samples. The original audio was recorded at a sampling rate of 12.5 kS/s with variable durations ranging from 0.3 to 0.8 seconds.

To interface with the antiferromagnetic MnIr device, the digital waveforms underwent specific conditioning steps. A 3rd-order Butterworth low-pass filter with 5 kHz cutoff frequency was applied to remove high-frequency noise components beyond the device operational bandwidth. Silence regions were trimmed using energy-based voice activity detection with a 16 dB threshold, employing a frame length of 128 samples and hop length of 4 samples to precisely identify speech segments. After trimming, all samples were zero-padded to a uniform length of 9890 samples to match the maximum duration in the dataset, ensuring consistent input dimensions across all recordings.

#### 2.2 Multi-Channel Physical Feature Extraction

The conditioned waveforms were then converted to analog voltage streams for direct input to the MnIr device array. The MnIr device array operates as a multi-channel nonlinear filter bank through bias-voltage-dependent resistance modulation. Feature diversity across channels is achieved by systematically varying three key device operating parameters: maximum electric field $V_{base} \in \{46, 48, 59, 64\}$ V (4 levels), DC bias field $V_{bias} \in \{-20, 46\}$ V (2 levels), and input voltage projection range $A \in \{[-20, 120], [-90, 50]\}$ V (2 configurations). Each unique combination of these parameters (4×2×2 = 16 total configurations) defines a distinct processing channel that constrains the device to evolve along a specific, history-dependent minor hysteresis loop. For experiments evaluating channel scaling effects in Figure 3d, we systematically selected subsets of 1, 8, 16, or 32 channels from the full parameter space to quantify the relationship between feature dimensionality and classification accuracy.

The signal processing pipeline for each channel proceeds through four stages. First, the normalized input signal is linearly mapped to the channel-specific voltage range $A = [V_{min}, V_{max}]$ through amplitude scaling. Second, the DC offset $E_{bias}$ is superimposed onto the projected signal to establish the operating point on the device hysteresis loop. Third, the combined voltage drives the MnIr device through its magnetic hysteresis loop, with the resistance state governed by the dynamic evolution function $R_{out} = f(R_{current}, E_{in}, E_{base}, E_{bias})$, where $R_{current}$ embodies the current resistance state encoding magnetic domain history, $E_{in}$ represents the instantaneous analog input voltage, $E_{bias}$ denotes the channel-specific DC offset, and $E_{base}$ refers to the device-specific switching threshold that defines the initial hysteresis state. Finally, the resulting resistance trajectory is downsampled by a factor of 5 to yield 1978 time points per channel, reducing computational overhead while preserving essential temporal features. The multi-channel outputs are stacked to form the final feature tensor with dimensions [$N_{channels}$ × 1978], which serves as input to the digital classification backend.

**2.3 Comparative Network Architectures**

We systematically evaluated eight distinct architectures to quantify the benefits of MnIr physical feature extraction, as shown in Figure 3d. To establish performance baselines without physical enhancement, we first tested two purely digital approaches and compared them directly with our single-channel physical preprocessing. Architecture 1 implemented a direct linear mapping from the flattened 9890-sample raw input to 10 output classes. With approximately 99,000 parameters, this linear baseline achieved only 23.4% accuracy. In contrast, when the raw signal was preprocessed by just a single physical channel of our MnIr device, a corresponding linear classifier (incorporating only 19,792 parameters) achieved an accuracy of 54.6% (Architecture 3). This direct comparison reveals that physical preprocessing more than doubles the baseline performance using a 5-times smaller backend classifier, confirming that the device possesses a powerful intrinsic nonlinear feature extraction capability. To further quantify the physical contribution, Architecture 2 and Architecture 4 compared the raw audio and the single-channel preprocessed signal using a shallow single-layer 1D convolutional neural network (CNN) backend (kernel size 3, 32 output channels). Because the learnable parameters of 1D convolutions depend exclusively on the number of input/output channels and are entirely independent of the temporal sequence length (STEP), both software networks possess the exact same low capacity of exactly 524 parameters. When processing the raw signal, this 524-parameter CNN reaches only 73.2% accuracy. However, when processing the device-output signal, the accuracy jumps dramatically to 92.8%, even though the device intrinsically downsamples the temporal data by a factor of 5 (from 9890 to 1978 time steps). Since the classification capacity is identical, this quantitative leap serves as definitive proof that the device's intrinsic physical nonlinearity dominates the feature extraction process.

The physics-enhanced architectures (Architectures 5–7) utilized MnIr device arrays for physical preprocessing with varying channel counts. Systematically increasing the physical channel count to 8, 16, and 32 channels yielded accuracies of 95.0%, 97.0%, and 97.2%, respectively. As the physical channels scaled, the digital parameter overhead of the single-layer CNN scaled linearly and remained extraordinarily lightweight—requiring only 1,210, 1,994, and 3,562 parameters, respectively. This scaling behavior confirms that expanding the nonlinear filter bank enriches feature diversity through increased parameter space coverage, with diminishing returns observed beyond 32 channels as the feature space approaches saturation and redundancy begins to dominate.

The ultimate system (Architecture 8) combines 32-channel physical feature extraction with a compact two-layer CNN backend to achieve optimal performance. The network accepts the 32-channel × 1,978-point feature tensor from the MnIr array and processes it through an initial batch normalization to stabilize the multi-channel input distribution. A first 1D convolutional layer expands the representation (kernel size 3, 32 channels), followed by batch normalization, tanh activation, and max pooling (stride 8). This is followed by a second 1D convolutional layer with identical channel dimensions, global average pooling to collapse the temporal dimension, and a final fully connected layer mapping to 10 output classes. The entire ultimate model contains precisely 6,730 trainable parameters and is trained using the AdamW optimizer with OneCycleLR scheduling (max learning rate 0.05, weight decay 0.001) over 1,000 epochs. This configuration achieves 99.8% accuracy, representing optimal synergy between physical preprocessing and backend refinement while maintaining much less parameter efficiency compared to conventional deep learning models

**2.4 Training Protocol and Statistical Validation**

For each experimental run, the 500-sample dataset was partitioned using stratified sampling to ensure balanced class representation in both training and test sets. Specifically, 5 samples per class were held out for testing (50 samples total), with the remaining 45 samples per class allocated to training (450 samples total). No separate validation set was employed; instead, model performance was evaluated directly on the held-out test set after training convergence. This approach maximizes the training data available for the small dataset while providing unbiased test set evaluation. To validate generalization capability and eliminate bias from specific train-test partitions, we performed 10 independent training runs with different random seeds controlling the stratified split. This Monte Carlo sampling approach generates a distribution of test accuracies across different data partitions, enabling calculation of mean performance and confidence intervals that reflect true generalization rather than overfitting to a single favorable split.

The training dynamics for the ultimate system demonstrate robust convergence across all experimental runs. As shown in Figure S12, the training loss drops rapidly from an initial value of 2.3 to approximately 1.0 within the first 50 epochs, indicating quick adaptation to the physical features. An intermediate plateau spanning epochs 50–150 sees test accuracy gradually rise from 86% to 96% as the model learns increasingly subtle discriminative patterns in the resistance trajectory space. A final refinement stage from epoch 150 onward shows performance stabilizing at the maximum achievable level, with test accuracy fluctuating minimally around 98%/100%. The absence of significant divergence between training and test metrics indicates effective regularization through the compact architecture design, with implicit dropout effects arising from physical noise in the MnIr devices providing additional regularization without requiring explicit dropout layers.

Across 10 independent runs with different train-test splits, the system achieved a mean test accuracy of 99.8%, confirming robust generalization and minimal sensitivity to specific data partitioning. Confusion matrix analysis reveals that the few remaining errors concentrate between acoustically similar digit pairs, particularly “one” versus “nine” which share similar vowel formants and temporal envelopes. Critically, digits that serve as mission-critical commands such as “zero” (abort signal) and “one” (initiate sequence) achieved 100% recognition accuracy across all test partitions, validating system reliability for safety-critical applications in extreme environments. This near-perfect performance, achieved with only thousands of trainable parameters and 16-32 physical preprocessing channels, demonstrates the power of physics-based feature extraction for enabling highly parameter-efficient neural architectures.

**Note S3：All-Hardware implementation**

The non-volatile multi-level resistance states of MnIr devices serve as analog synaptic weights for implementing both convolutional and fully-connected layers in the neural network backend. Unlike volatile reservoir states that require continuous input stimulation, these weight resistances remain stable without power supply, enabling direct in-memory computing where stored weights simultaneously perform computation. We experimentally characterize the accessible resistance distribution of our MnIr arrays under controlled programming protocols, obtaining a discrete set of achievable conductance levels. These values, rather than idealized uniform quantization bins, form the physical constraint set for weight encoding, naturally capturing device-to-device variations and programming nonidealities intrinsic to the antiferromagnetic switching process.

To bridge the gap between continuous-valued software weights and discrete hardware conductances, we employ non-uniform quantization-aware training where network weights are constrained during training to match the experimentally accessible resistance levels. During the forward pass, each continuous weight $w$ is mapped to its nearest quantized value, ensuring all trained weights correspond to physically realizable device states. For backpropagation, we apply the straight-through estimator, allowing gradients to flow through the quantization function as if it were an identity operation. This approach enables standard gradient descent optimization while respecting hardware constraints: the network learns to position weights at locations where quantization-induced errors minimally impact task performance. Training proceeds iteratively over multiple epochs, with validation accuracy progressively improving as the network adapts its weight distribution to accommodate the non-uniform quantization imposed by the physical device characteristics, ultimately achieving 99.8% recognition accuracy when the 32-channel physical features are combined with the two-layer quantized CNN backend (detailed training curves in Figure S13).

Convolutions and fully-connected operations both reduce to matrix-vector multiplications, directly implementable through Ohm's law on crossbar arrays. Each MnIr device programmed to conductance $G_{ij}$ is positioned at the crosspoint of row $i$ and column $j$, with input voltages $V_j$ applied along columns and output currents $I_i = \sum_j G_{ij} V_j$ collected from rows, physically realizing the multiply-accumulate operations in a single time step with O(1) complexity (Figures S8-S11). For convolutional layers, we employ the im2col transformation to unfold image patches into column vectors, allowing 2D convolutions to be executed as standard matrix multiplications on the crossbar. Negative weights are handled using differential pairs where each synapse consists of two devices ($G^+, G^-$) encoding $w_{ij} = G_{ij}^+ - G_{ij}^-$, with outputs subtracted in the analog current domain. This fully analog in-memory computing architecture eliminates the von Neumann bottleneck, achieving energy efficiency superior to digital implementations while the intrinsic radiation hardness and thermal stability of antiferromagnetic materials ensure reliable operation across the entire 1.6-500 K temperature range and under 55 T magnetic fields without performance degradation.

## Note S4: Demonstration of capability in extreme aerospace environments

To demonstrate the practical capability of the device in extreme aerospace environments, we utilize the device's self-refreshing property and strong nonlinearity to post-process gyroscope angular velocity data from the Jupiter Orbiter (Juno), extracting low-noise angular velocity signals from a high-noise environment.

### 4.1 Extraction of Juno flight data based on the NASA SPICE system

To construct a high-fidelity angular velocity benchmark dataset, we perform data extraction based on the SPICE Toolkit and relevant Kernels files provided by the NASA Navigation and Ancillary Information Facility (NAIF).[1] We select flight data from the Juno spacecraft [2] near the perijoves (PJ) as the subject of study. Specifically, the data span covers the complete orbital periods from PJ45 to PJ63. Using core functions of the SPICE system, such as spkezr and subpnt, we extract the spacecraft's precise position, velocity, and angular velocity relative to Jupiter. These kinematic parameters, calculated based on official ephemerides and post-processed by ground stations, represent the spacecraft's true trajectory under ideal conditions and serve as the ground truth reference values ($\boldsymbol{\omega}_{gt}(t)$) for subsequent training and validation of the noise reduction effect.

Meanwhile, to better analyze the spacecraft's motion characteristics within Jupiter's body-fixed coordinate system, we perform coordinate system transformations and trajectory solutions. Raw SPICE data is typically based on the J2000 inertial frame; we transform this into the JuFrame coordinate system using rotation matrices. In this frame, we precisely calculate Juno's 3D spatial position coordinates during the aforementioned perijove passes.

### 4.2 Calculation of Jupiter's extreme magnetic field distribution and simulated noise construction

At the perijove, the intensely varying magnetic field interferes with the gyroscope's angular velocity measurements. To simulate this interference, we first calculate the magnetic field intensity and vector distribution experienced by the spacecraft along its orbit, combining Juno's real-time position in the JuFrame system with Jupiter's intrinsic magnetic field model JRM33 [3]. The noise signal is obtained via the following equations:

$$\boldsymbol{\omega}_{corrupted}(t) = \boldsymbol{\omega}_{gt}(t) + \boldsymbol{n}_{mag}(t)$$

$$\boldsymbol{n}_{mag}(t) = f(\boldsymbol{B}(t)) \approx Coff_{mag} \cdot \frac{d\boldsymbol{B}(t)}{dt}$$

where $Coff_{mag} = \begin{bmatrix} 5\times10^{-7} & 1\times10^{-8} & 1\times10^{-8} \\ 1\times10^{-8} & 5\times10^{-7} & 1\times10^{-8} \\ 1\times10^{-8} & 1\times10^{-8} & 5\times10^{-7} \end{bmatrix}$ is the custom magnetic sensitivity coupling coefficient.

### 4.3 Angular velocity denoising

To recover the true spacecraft motion information from the magnetically interfered gyroscope data, we construct a physics-based linear correction model. The core idea of the algorithm is to use magnetic field data as prior knowledge to inversely cancel out the magnetic coupling noise.

The input feature space of the model is defined as 6-dimensional. For any given time $t$, we construct a joint input vector $\boldsymbol{X}(t)$ by concatenating the corrupted angular velocity $\boldsymbol{\omega}_{corrupted}(t)$ measured by the gyroscope and the calculated environmental magnetic field vector $\boldsymbol{B}(t)$:

$$\boldsymbol{X}(t) = \begin{bmatrix} \boldsymbol{\omega}_{corrupted}(t) \\ \boldsymbol{B}(t) \end{bmatrix}$$

The denoising task aims to find a mapping relationship to reconstruct the 3-dimensional estimated true angular velocity $\hat{\boldsymbol{\omega}}(t)$ from this 6-dimensional input. This process can be expressed as:

$$\hat{\boldsymbol{\omega}}(t) = W \cdot \boldsymbol{X}(t) + b$$

Here, $W \in \mathbb{R}^{3\times6}$ is the weight matrix to be solved, which contains both the retention weights for the original signal and the compensation weights for the magnetic field interference terms.

We construct a dataset using the processed data from 19 perijove orbits. The dataset is divided into a training set and a testing set: time-series data from selected orbits forms the training set to learn the mapping between the magnetic field and the noise; the remaining orbital data serves as the testing set to verify the denoising effect. In the training phase, the algorithm uses the ground truth angular velocity $\boldsymbol{\omega}_{gt}(t)$ extracted from the SPICE system as the supervision target. The parameter matrix $W$ is optimally estimated via linear regression, thereby achieving efficient separation of magnetic interference and signal restoration in the testing set.

Figure S15 displays the results using the 48th perijove (PJ48) as the test set. Compared to the original noisy signal, the noise in the post-processed angular velocity was reduced by 41.46%. The processed angular velocity can improve the precision of Juno's real-time attitude correction. This preliminarily demonstrates the device's capability to assist sensors in data processing within extreme space environments.

**Note S5: Three typical recognition tasks**

To simulate the real-time detection of astronaut activities and environmental perception within a space capsule, we utilized the device to perform a three-fold recognition task comprising visual object, action, and gesture recognition.

**5.1 Object Recognition**

In Figure 4d, we utilize the Caltech-101 dataset for visual object recognition. To evaluate the robustness of the device under low-quality imaging conditions, we expand the original dataset by generating data with reduced contrast and introduced Gaussian noise (specific images are shown in Figure S17). We employ a linear transformation method to attenuate the image contrast. For an input image $I$, the corresponding low-contrast image $I_{low}$ is generated via the following formula:

$$I_{low}(x,y) = \text{clip}(\alpha \cdot I(x,y) + \beta, 0{,}255)$$

where $I(x,y)$ represents the pixel value of the original image at coordinates $(x,y)$, and $\alpha$ is the contrast control factor. When $0 < \alpha < 1$, the image contrast is reduced. $\beta$ denotes the brightness bias. In this experiment, we set $\alpha = 0.6$ and $\beta = 0$. The clip function is used to truncate the calculated pixel values within the valid grayscale range $[0{,}255]$ to prevent numerical overflow.

We select 10 representative biological categories (including animals and plants: Ant (O1), Bonsai (O2), Butterfly (O3), Crab (O4), Dalmatian (O5), Dolphin (O6), Elephant (O7), Kangaroo (O8), Lotus (O9), and Sunflower (O10) from the Caltech-101 benchmark. For each category, 10 base images were collected. Simultaneously, we generate low-contrast and noisy versions for each image, constituting a dataset of 300 images. Each image has a resolution of 100×100 pixels, with the test set accounting for 20% of the dataset. After physical feature mapping by the MnIr device, the hardware-implemented classifier achieved an optimal classification accuracy of 100%.

To evaluate the versatility of the device for biometric identification (such as astronaut identity authentication), we also validated the exact same computing pipeline on human face recognition using the ORL Database of Faces under identical degradation conditions (https://www.kaggle.com/kasikrit/att-database-of-faces). The proposed all-analog computing system similarly achieved 100% recognition accuracy across 10 individuals. To respect individual privacy rights and adhere to open-access publishing guidelines regrading the republication of identifiable personal portraits without explicit written consent, the Caltech-101 object dataset is adopted as the primary visual demonstration in the manuscript.

### 5.2 Action Recognition

In Figure 4b, the UTD-MHAD (University of Texas at Dallas Multimodal Human Action Dataset) is adopted as the benchmark for the action recognition task. This dataset is acquired using a Microsoft Kinect sensor, providing 3D skeleton sequences containing 20 joint points at a frame rate of approximately 30 fps.

In this experiment, we select eight typical action classes for evaluation: Right arm swipe to left (A1), Draw circle clockwise (A2), Front lunge (A3), Cross arms in the chest (A4), Stand to sit (A5), Squat (A6), Pick up and throw (A7), and Jogging (A8). The dataset includes 8 subjects (4 males and 4 females). Following the cross-subject protocol, data from subjects 1, 2, 4, 5, 6, and 8 are used for training, while data from subjects 3 and 7 are reserved for testing.

We convert the raw skeleton sequences into a format suitable for device processing through the procedure described below, ultimately generating an 18×125 matrix to serve directly as the device input.

1. Skeleton Grouping

We divide the original 20 skeleton joints into 6 skeletal groups based on anatomical connections:

- Group 1 (Head and Torso): Head, shoulder center, spine, hip center;
- Group 2 (Left Arm): Left shoulder, left elbow, left wrist, left hand;
- Group 3 (Right Arm): Right shoulder, right elbow, right wrist, right hand;
- Group 4 (Left Leg): Left hip, left knee, left ankle, left foot;
- Group 5 (Upper Right Leg): Right hip, right knee;
- Group 6 (Lower Right Leg): Right ankle, right foot.

For each skeletal group, we obtain the centroid coordinates $(x, y, z)$ by calculating the arithmetic mean of the coordinates of all joints within the group. This process compresses the original 60-dimensional spatial information (20 points × 3 coordinates) into 18 feature channels (6 skeletal groups × 3 coordinate axes).

2. Temporal Normalization

Due to the natural variation in the duration of different action samples, we define a fixed time window. The temporal length of all action sequences is uniformly normalized to $T =$ 125 frames. Sequences shorter than 125 frames are zero-padded at the end, while sequences exceeding 125 frames are truncated.

3. Signal Encoding

To adapt the physical signals for direct device input, we independently standardize and scale each of the 18 feature channels. Specifically, we first apply Z-score standardization to the data, followed by encoding using the hyperbolic tangent (Tanh) activation function:

$$x'_{t,i} = \tanh(0.6 \times \frac{x_{t,i} - \mu_i}{\sigma_i})$$

where $x_{t,i}$ represents the original value of the $i$-th feature channel at time $t$, and $\mu_i$ and $\sigma_i$ are the mean and standard deviation of that channel, respectively. The coefficient 0.6 is used to adjust the data distribution to match the sensitive operating range of the device. The final output is an 18×125 (feature channels × time steps) spatiotemporal feature matrix.

**5.3 Gesture Recognition:**

We connect an MPU-6050 accelerometer to a self-made ESP-12F board to collect real-time acceleration data. To avoid the impact of gravitational acceleration (g ≈ 9.8 m/s$^2$) and better reflect the alternative in acceleration, the raw data collected by the accelerometer is processed by the ESP-12F chip to calculate the differential data. We have recorded a training and test dataset with a length of 20 groups and 10 groups, respectively, each group with 5 distinct gesture data samples corresponding to five states: Up (G1), Down (G2), Left (G3), Right (G4) and Stationary (G5). Before recording these gestures, we calibrate our sensor to prevent initial readings that might arise from thermal noise.

**Note S6: Parallel spatiotemporal coding strategy based on 2D sliding windows**

In conventional image recognition tasks utilizing memristors or other neuromorphic devices, a one-dimensional (1D) pixel-by-pixel input method is typically adopted.[4–6] While this approach leverages the memory characteristics of the devices, it disrupts the local spatial correlations within the image, and the processing speed is constrained by the total number of pixels. As the patterns recognized by the UAV in Figure 5 exhibit significant variations in pixel values but a limited variety of pixel types, the recognition performance primarily depends on the device's ability to extract local edge features (such as corners). Consequently, the 1D pixel-by-pixel input method is evidently limited in this scenario. This paper proposes a parallel spatiotemporal coding strategy based on a 2D sliding window. By segmenting the image into a sequence of local patches, this method achieves an organic integration of spatial feature extraction and the temporal dynamics (path-dependent mechanism) of the device.

In Figure 5, the ground patterns captured by the onboard camera are first converted to grayscale and resized to images with pixel values ranging from 0 to 255. Subsequently, the image is segmented into 100 local patches, each with a size of 10×10. This process involves encoding in two dimensions:

Spatial Dimension: The 10×10 pixels within each patch are mapped in parallel to 100 physical device nodes. This enables the device array to directly capture the local spatial features of the image.

Temporal Dimension: The sequence of patches, generated by scanning the image with the sliding window at a specific stride, is sequentially fed into the same device array. The resistivity of the MnIr layer depends not only on the current patch input but is also influenced by the residual state following the excitation of preceding patches.

The final feature matrix is obtained by downsampling this sequence of 100 patch states.

**Note S7: Comparison on power consumption, response frequency and environmental resilience**

**7.1 Comparison on power consumption:**

We compare this device with other neuromorphic devices that possess image recognition capabilities and have the potential for real-time interaction with UAVs. The comparison focuses on energy consumption, response frequency, and tolerance to extreme environments.

Given the variation in recognition tasks across different studies, the comparison of energy and response frequency excludes the influence of peripheral circuits. Instead, it is derived from the total energy consumed and the total time required for the device to perform a single write and read operation.

The energy consumed by the device for a single read-write operation ($E$) is calculated as:

$$E = E_W + E_R$$

$E_W$ is the energy for write operation and can be calculated as the elastic energy for switching the polarization in ferroelectric oxides:

$$E_W = 1/2 P_s V S$$

where $P_s$ is the saturation ferroelectric polarization (~25 μC/cm$^2$ for PMN-PT), $V$ is the switching voltage, and $S$ is the cell area. Considering future integration, it can be scaled down to a footprint area of 100×100 nm$^2$ with PMN-PT thickness of 1 μm for integrated device applications, the energy consumption for switching the polarization would be $E_W = E_{Switching} = 1/2 P_s V S = 1/2 \times 25\ \mu C/cm^2 \times 1.2\ kV/cm \times 1\ \mu m \times 100\ nm \times 100\ nm \approx 0.15$ fJ.[7–10]

$E_R$ is the energy for read operation and can be calculated as

$$E_R = I^2 \times R \times t$$

where $R$ represents the average resistance of the MnIr layer in this work, $I$ represents the read current and $t$ represents the current duration time, which could be shortened to 1 μs by pulse generator. $E_R = I^2 \times R \times t = (1\ \mu A)^2 \times \sim 50\ \Omega \times 1\ \mu s \approx 0.05$ fJ. Therefore, the energy consumed by the device for a single read-write operation can be: $E = E_W + E_R \approx 0.2$ fJ.

**7.2 Response frequency:**

The device responses to analog input and generates features, relying on nonvolatile path evolution rather than volatile decay, its operational timescale is decoupled from fixed physical relaxation time constants. As long as the input does not exceed the ultimate switching speed of the ferroelectric domains, the device can adapt to analog signals spanning different timescales. So the response frequency of the device ($f$) is:

$$f = \frac{1}{T} = 1\ THz$$

where $T$ is the ferroelectric domain switching time of approximately 1 ps.[11–13]

**7.3 Environmental resilience:**

To compare the tolerance to extreme environments, we adopt a scoring metric:

$$Score = \frac{1}{3} \times 100 \times (\frac{M}{M_max} + \frac{X}{X_max} + \frac{T}{T_max})$$

where $M$, $X$, and $T$ represent the compared device's maximum withstandable magnetic field intensity, maximum X-ray irradiation dose, and operating temperature range, respectively. The denominators correspond to the maximum values of these indicators among the compared devices. Similar to the comparison method used in Figure 3e, as the specific experimental values for these three indicators may not be reported in other literature, we used the theoretical limit values of the constituent materials of the devices as the basis for comparison.

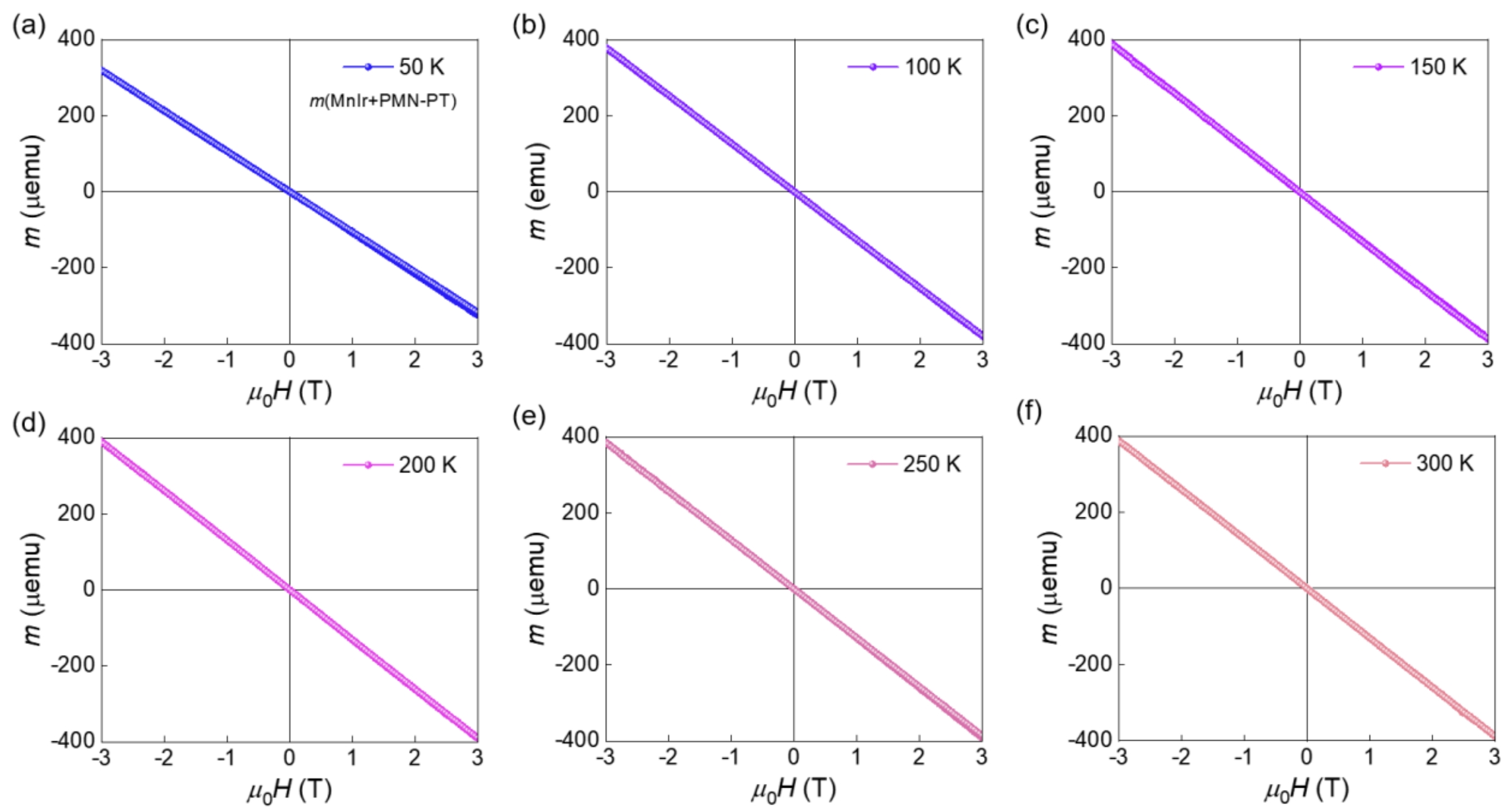

**Figure S1 |** Magnetic measurement results of MnIr/PMN-PT heterostructure from 50 to 300 K with in-plane magnetic fields.

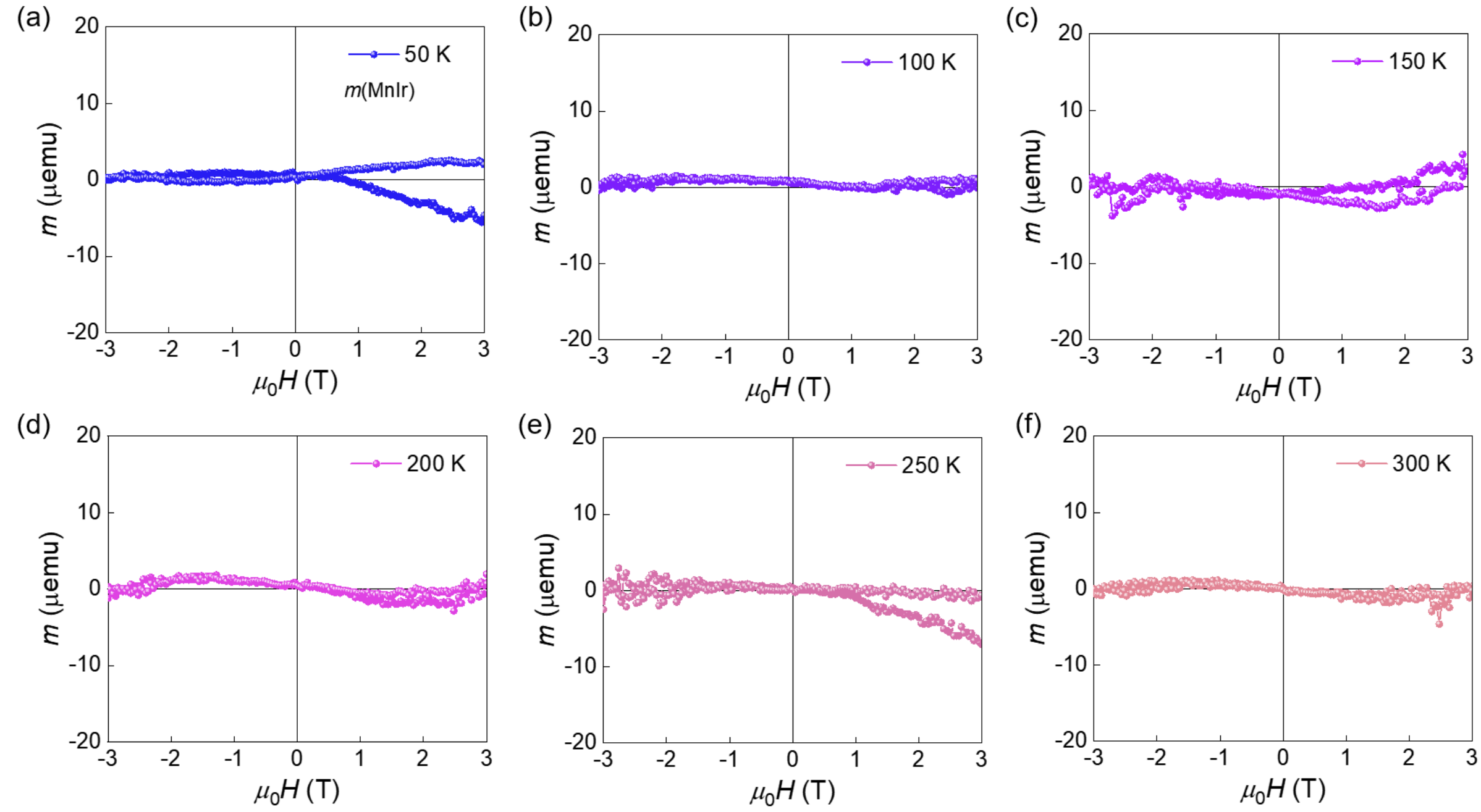

**Figure S2 |** Magnetic properties of MnIr thin film following the subtraction of the diamagnetic background presented in Figure S1.

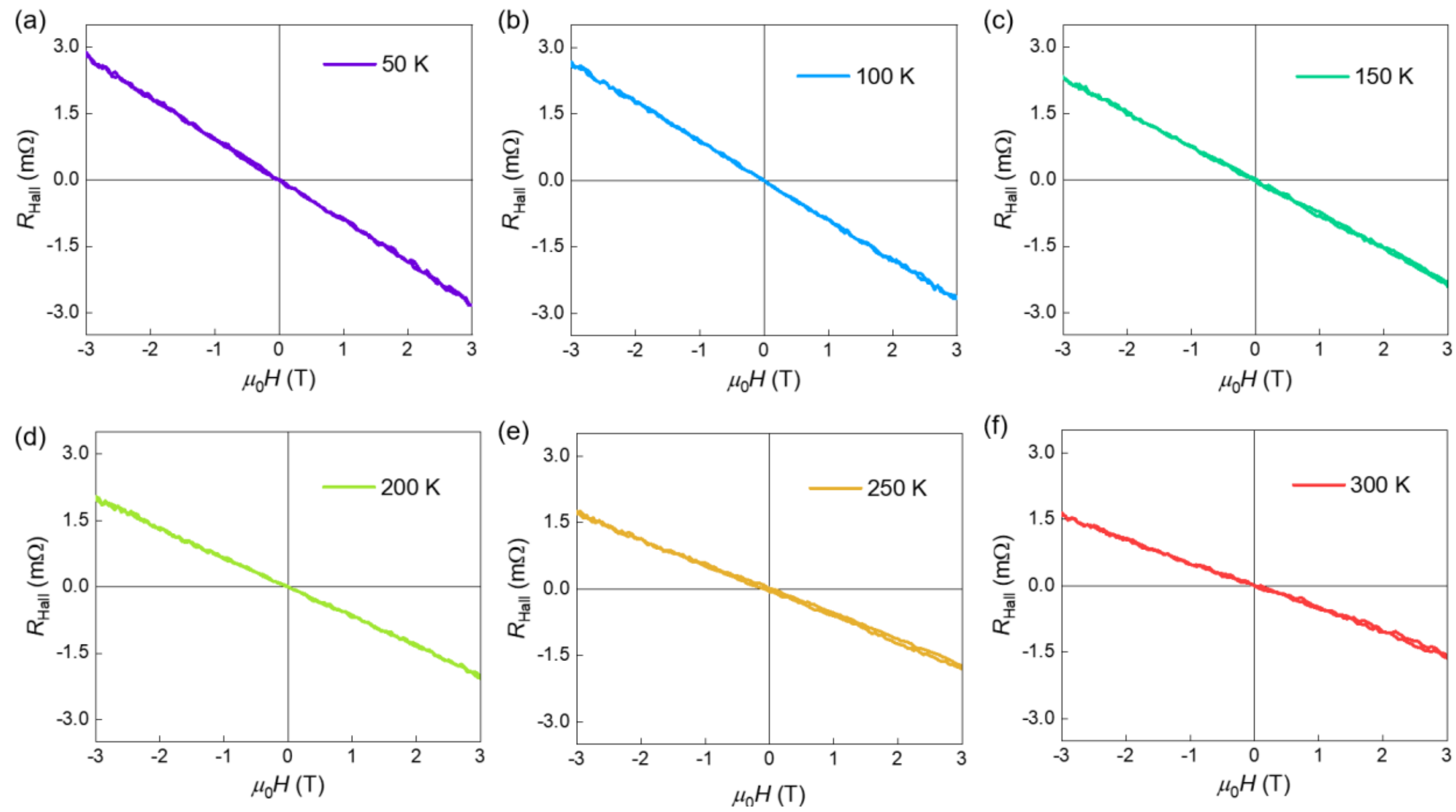

**Figure S3 |** Measurement results of Hall resistance for MnIr/PMN-PT heterostructure across the temperature range of 50–300 K.

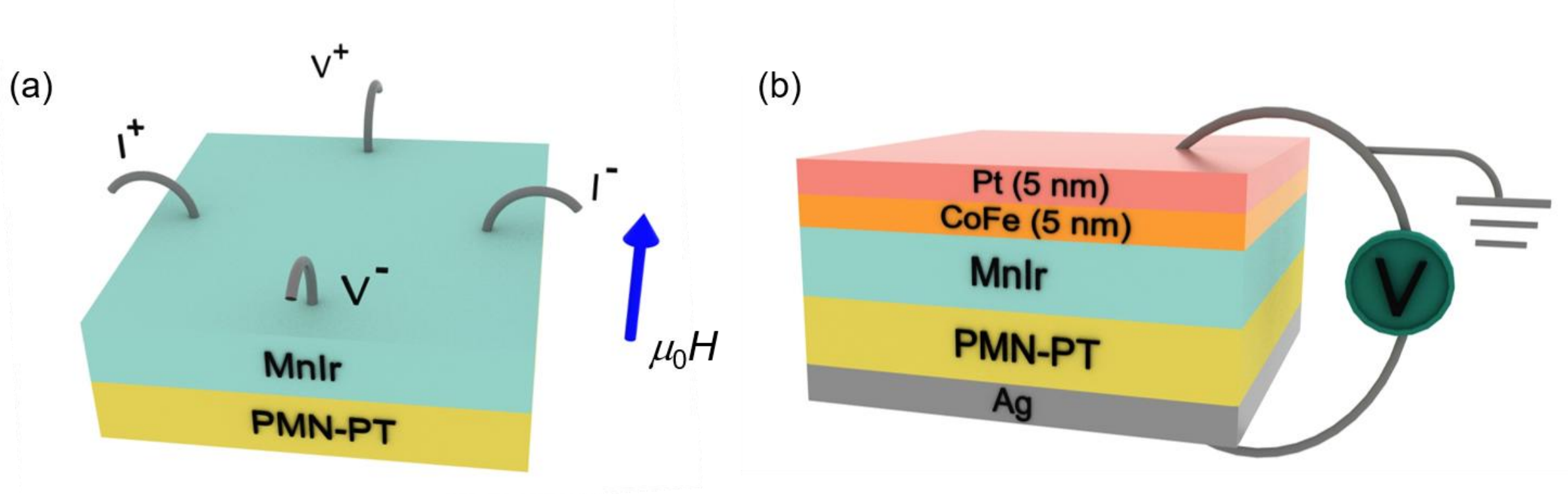

**Figure S4 |** (a) Schematic geometry of the Hall resistance measurement. (b) Sketch of the Pt/$Co_{90}Fe_{10}$/MnIr/PMN-PT heterostructure for magnetic measurement under electric fields.

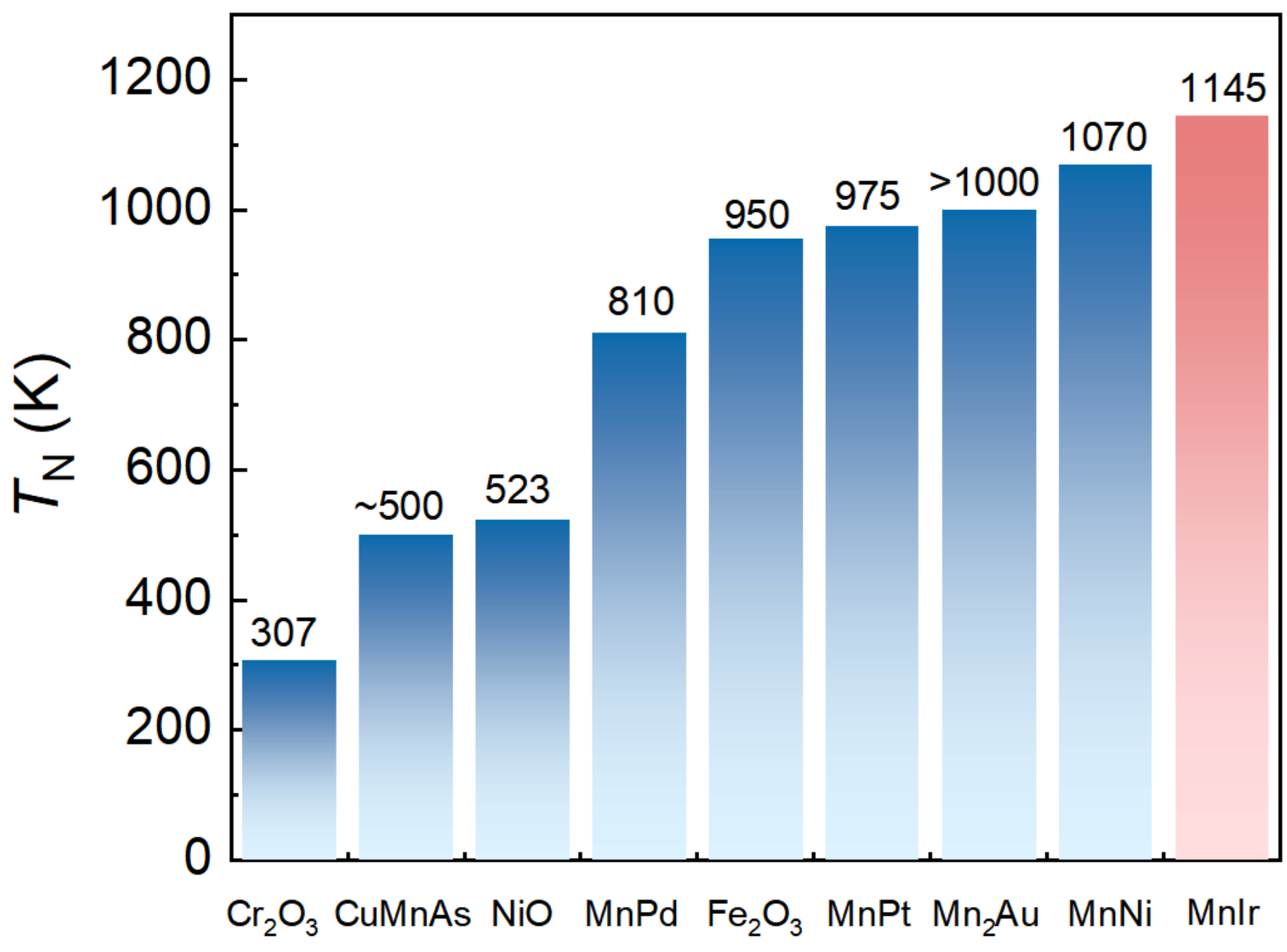

**Figure S5 |** Comparison of Néel temperatures for representative antiferromagnetic materials.

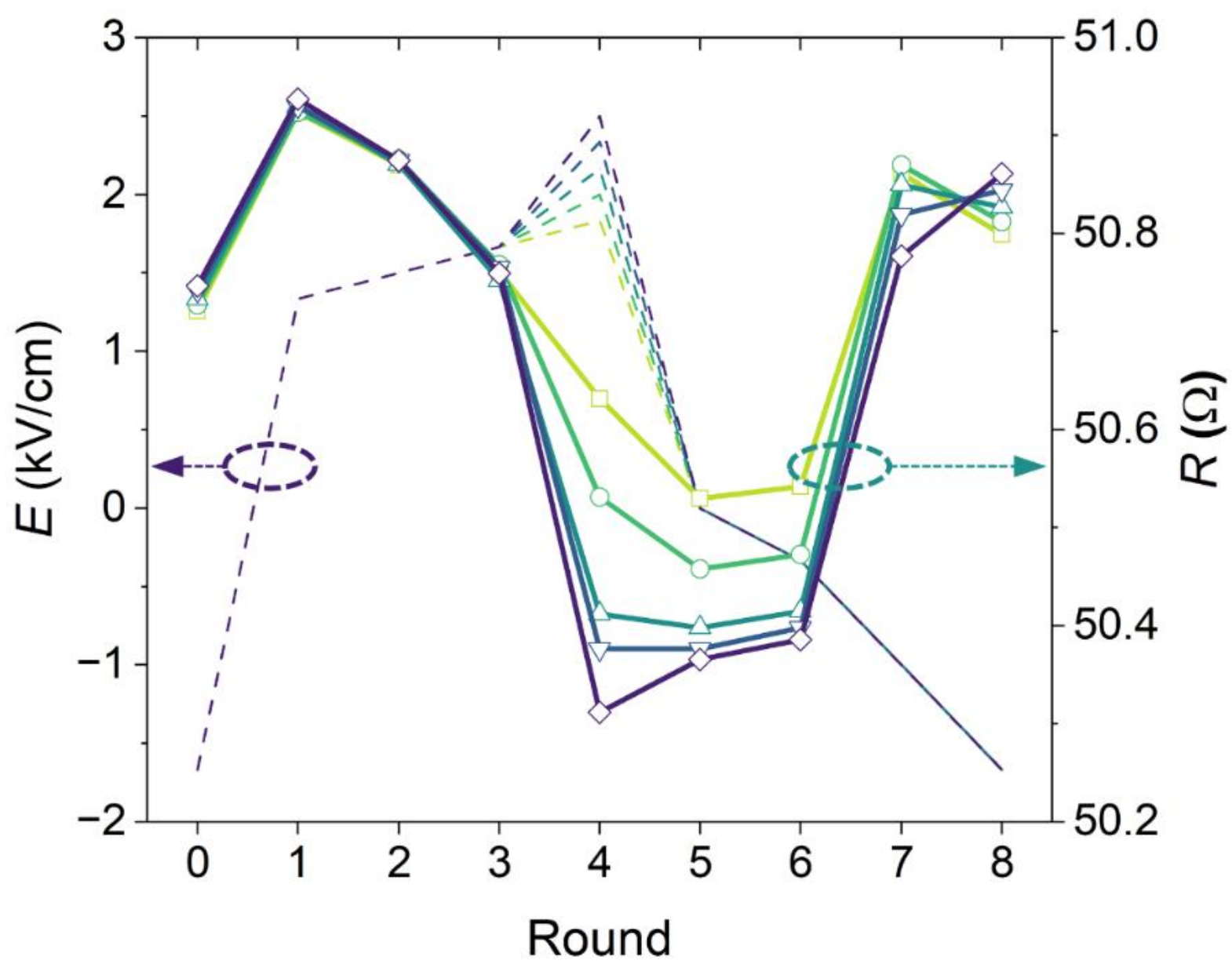

**Figure S6 |** Transformation of temporal information into spatial resistance states via strain-mediated hysteresis.

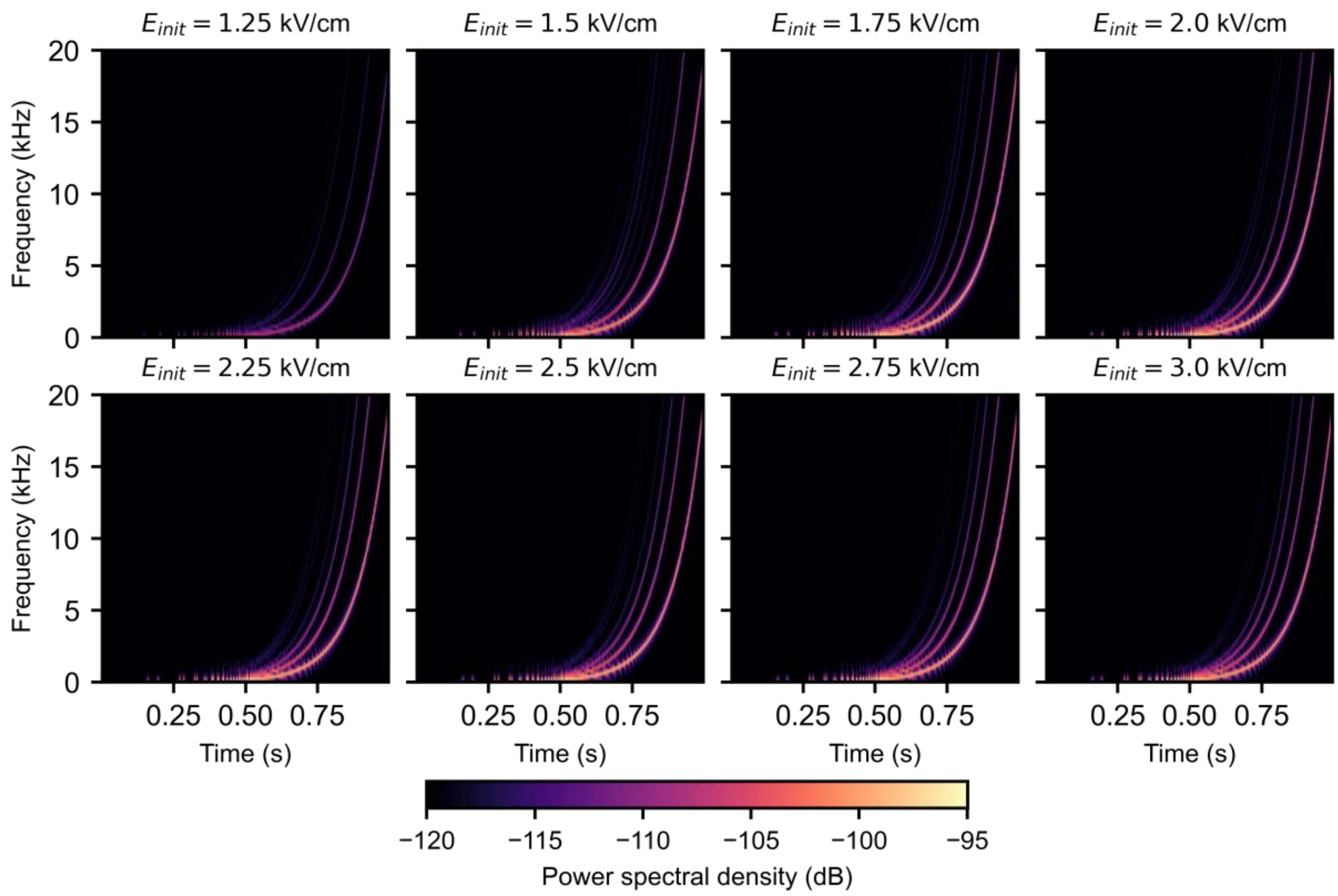

**Figure S7 |** Frequency-domain harmonic analysis revealing tunable nonlinear dynamics in the MnIr/PMN-PT heterostructure. Short-Time Fourier Transform (STFT) spectrograms of the device output responses under different initial polarization states. To quantitatively assess the physical nonlinearity, the PMN-PT substrate is excited by an exponential chirp electric field $E_{in}(t) = E_{bias} + E_{amp}\sin(\phi(t))$, where the instantaneous frequency $f(t) = \frac{1}{2\pi}\frac{d\phi(t)}{dt}$ sweeps logarithmically from $f_0 = 1$ Hz to $f_1 = 20$ kHz over a duration of $T = 1.0$ s. The amplitude parameters are set to $E_{bias} = 0$ kV/cm and $E_{amp} = 1$ kV/cm. Prior to the chirp excitation, the device is initialized to distinct segments of the hysteresis loop by applying different positive polarization electric field $E_{init}$ ranging from 1.25 kV/cm to 3 kV/cm (as indicated above each panel). The dynamic resistance evolution $R(t)$ is continuously monitored via a constant probe current ($I_{read} = 1$ mA), and the power spectral density (PSD, in dB) is calculated. Alongside the fundamental frequency trace, prominent higher-order harmonics (2nd, 3rd, 4th, etc.) are clearly generated. These distortion products originate from the strain-mediated nonlinear transfer function of the antiferromagnetic domains. The energy distribution and the emergence of specific high-order harmonic branches evolve with $E_{init}$. This frequency-domain evidence validates that the intrinsic physical nonlinearity is tunable, serving as the hardware foundation for constructing the heterogeneous nonlinear filter bank used in our preprocessing-free computing architecture.

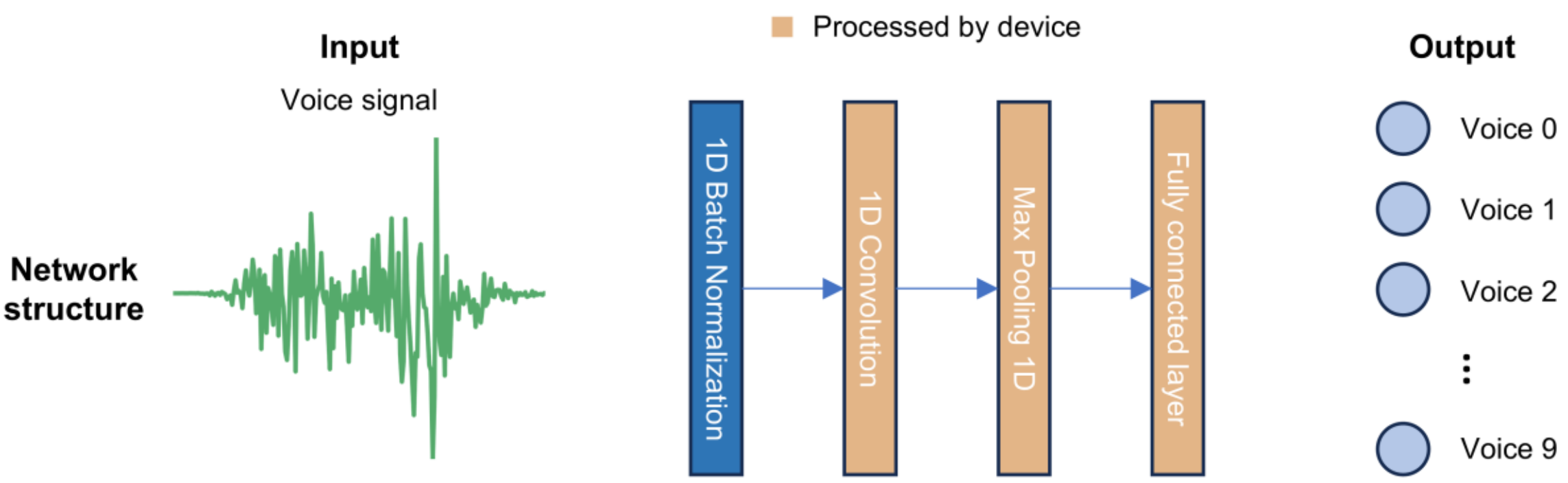

**Figure S8 |** Schematic diagram of single-layer CNN architecture.

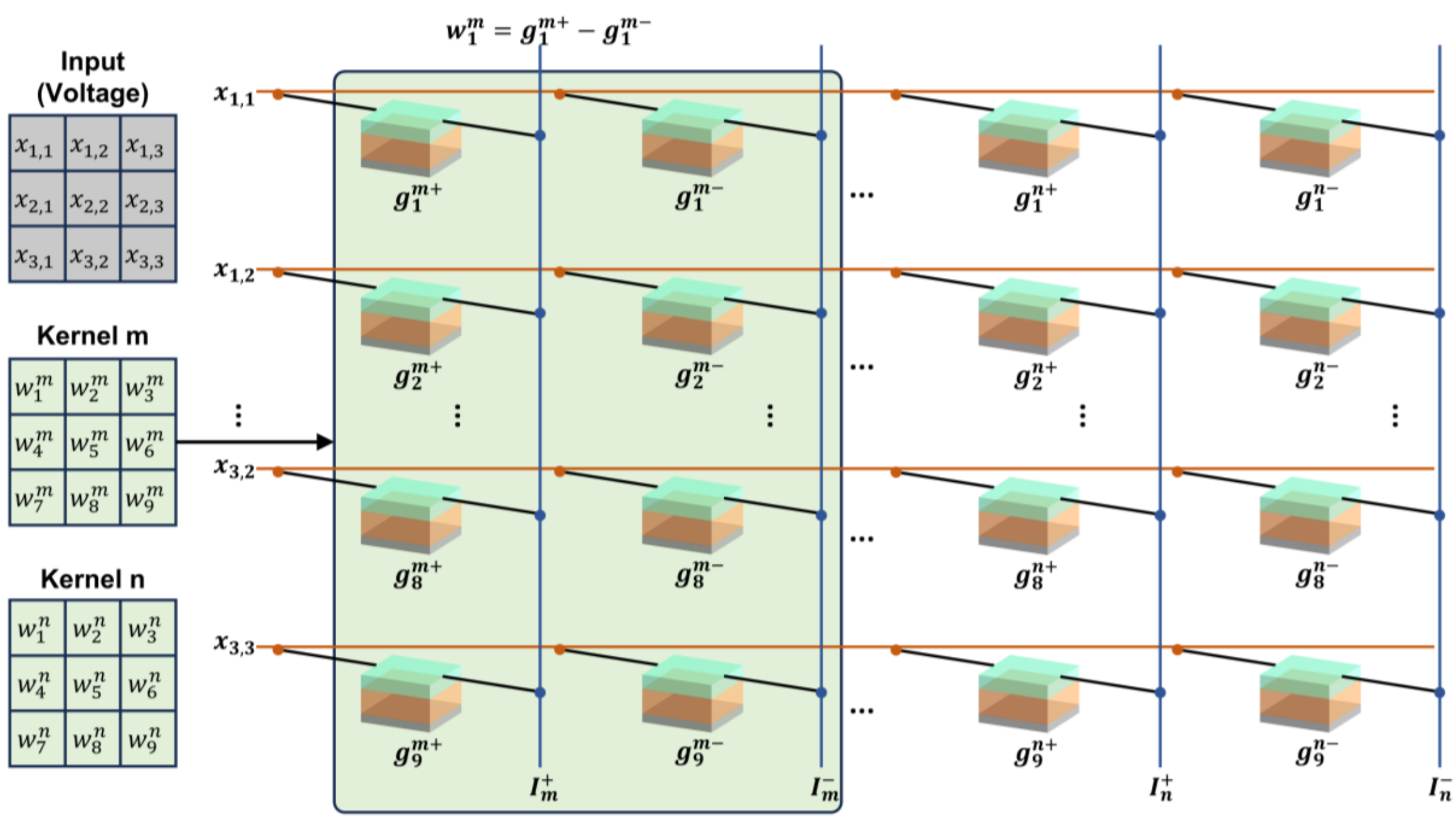

**Figure S9 |** Schematic diagram of performing convolution using the devices.

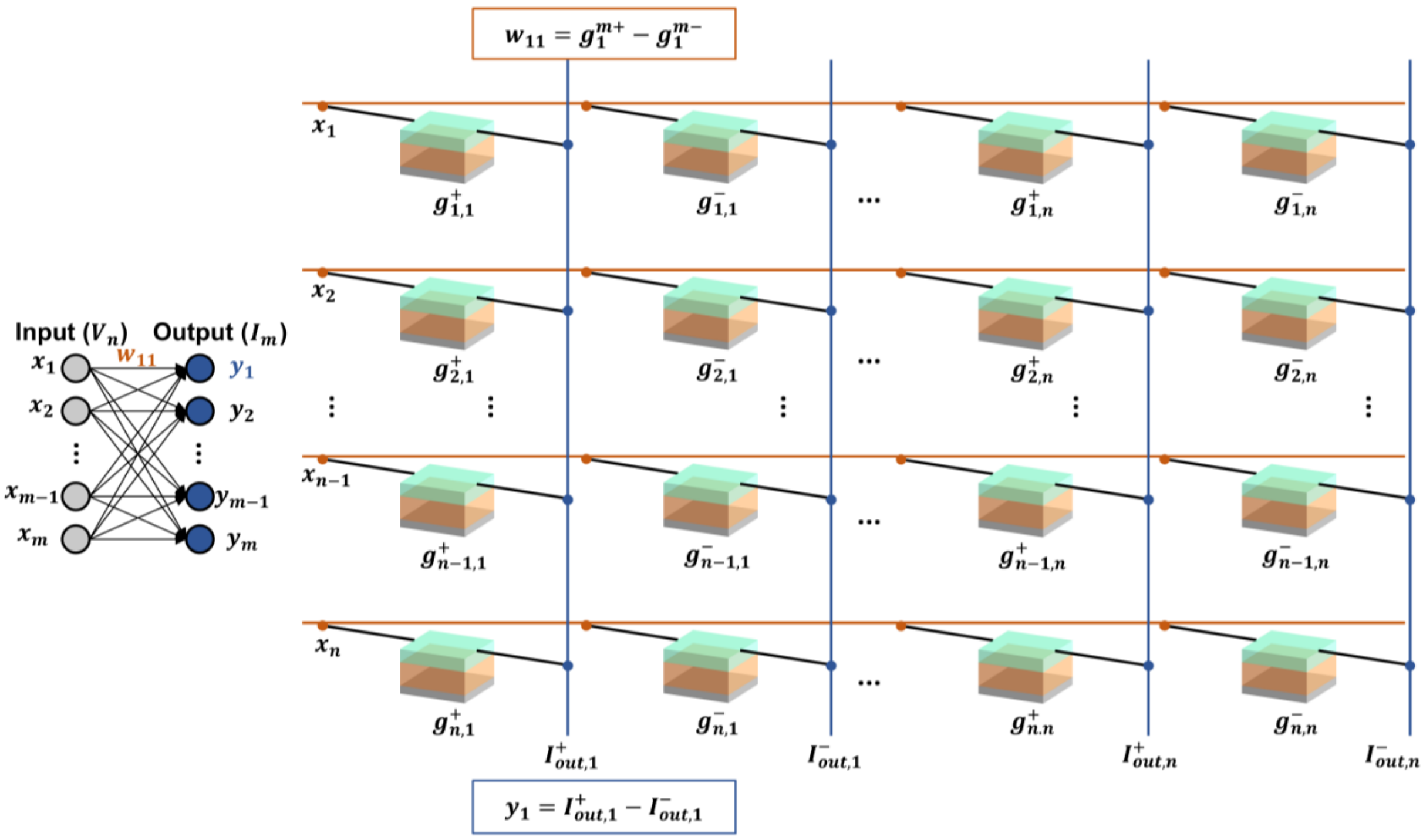

**Figure S10 |** Schematic diagram of performing a fully connected layer using the devices.

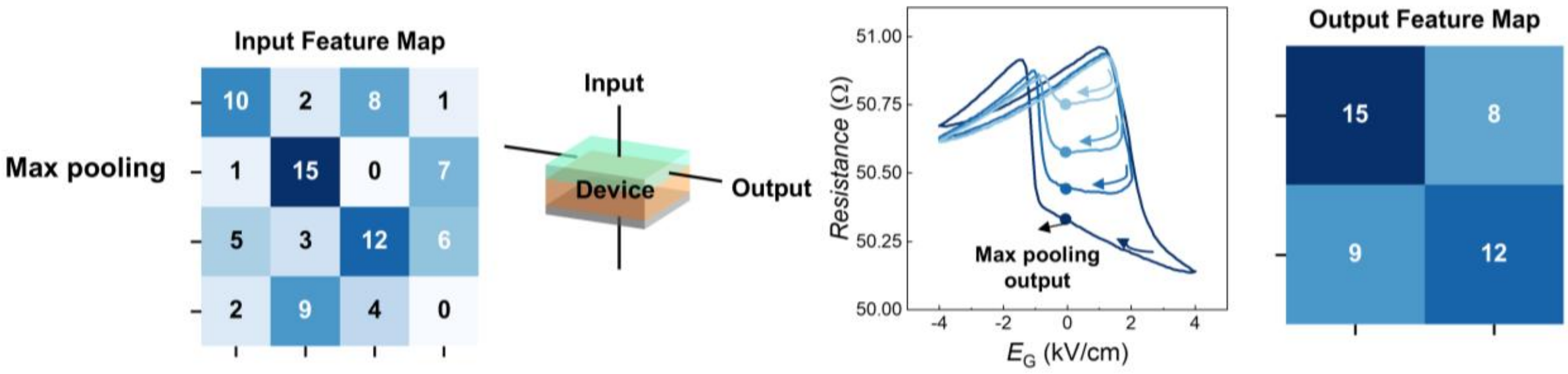

**Figure S11 |** Schematic diagram of performing max pooling using the devices.

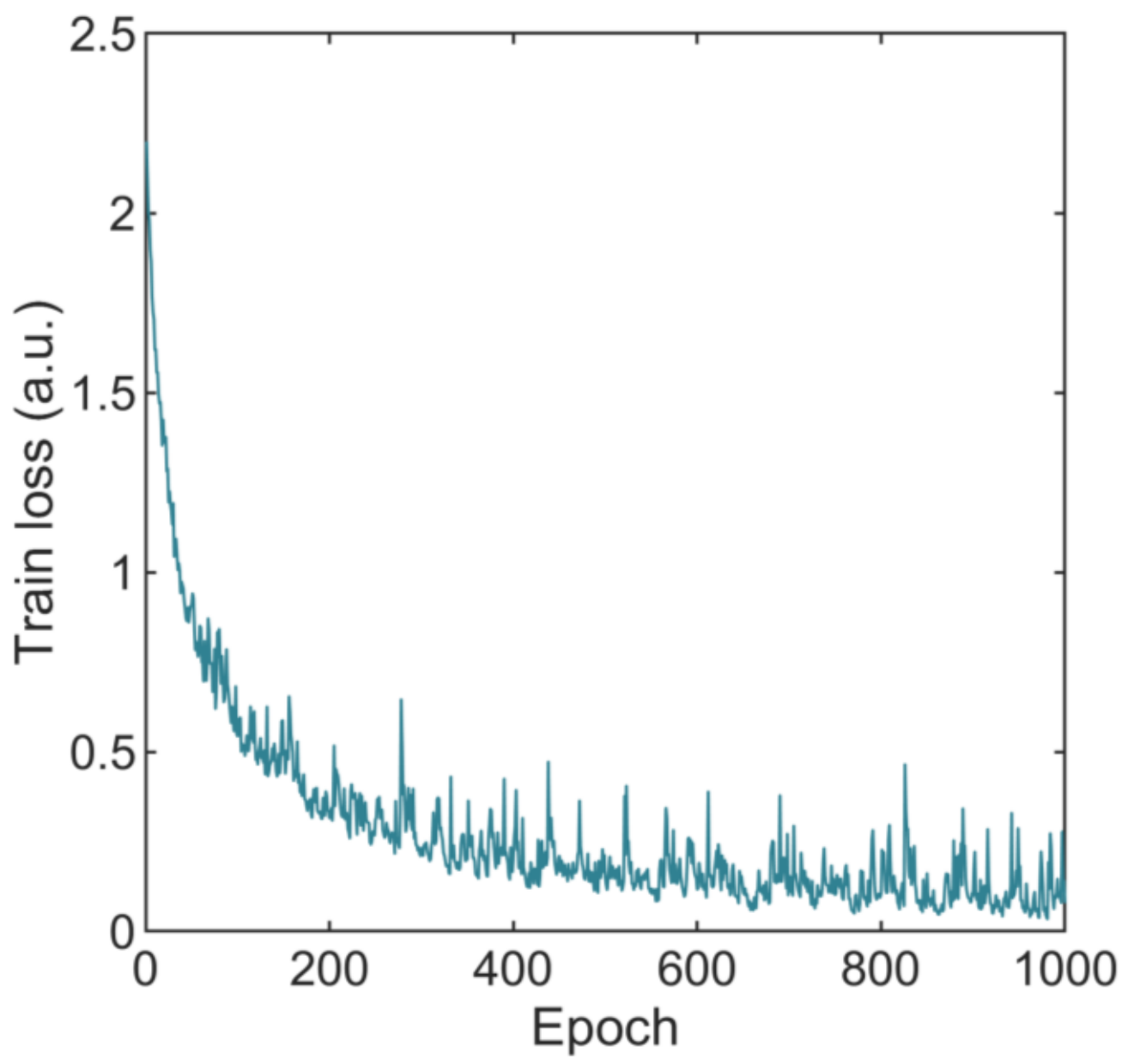

**Figure S12 |** Train loss of two-layer CNN model with full-precision.

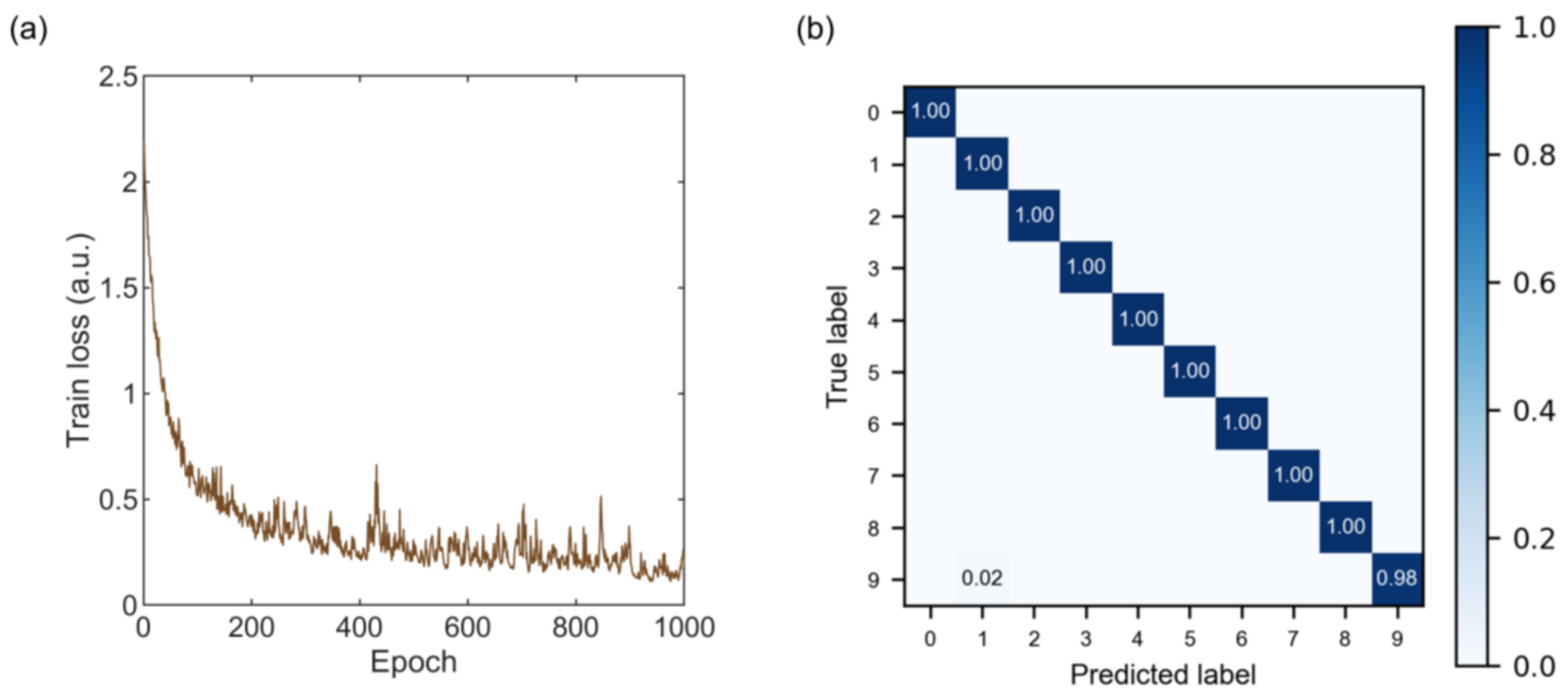

**Figure S13 |** Train loss of two-layer CNN model implemented by device array (a) and the confusion matrix of ten-fold cross-validation (b).

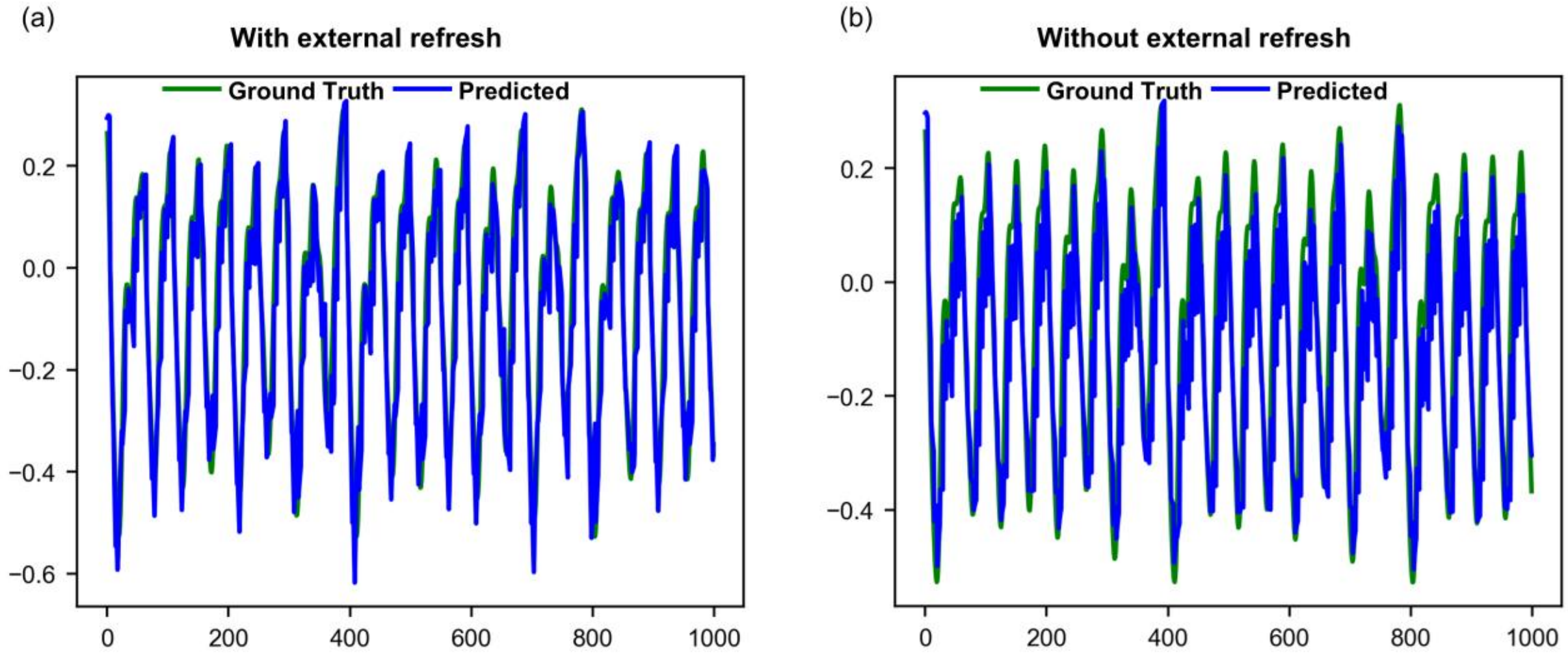

**Figure S14 |** Mackey-Glass (MG) sequence prediction task[28]. (a) The MSE of MG sequence 5 steps prediction is 0.00768 with external refresh. (b) The MSE of MG sequence prediction is 0.01623 without external refresh.

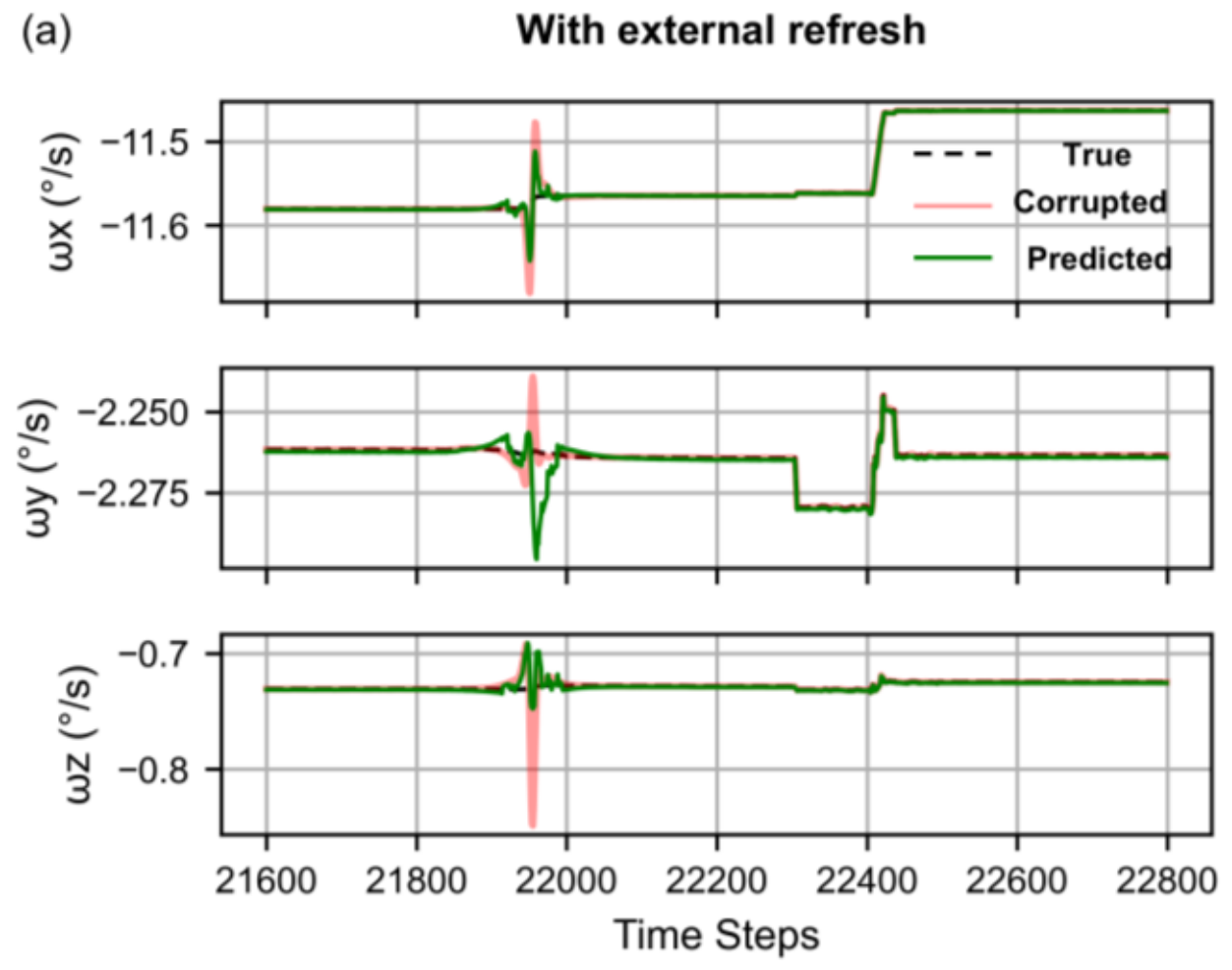

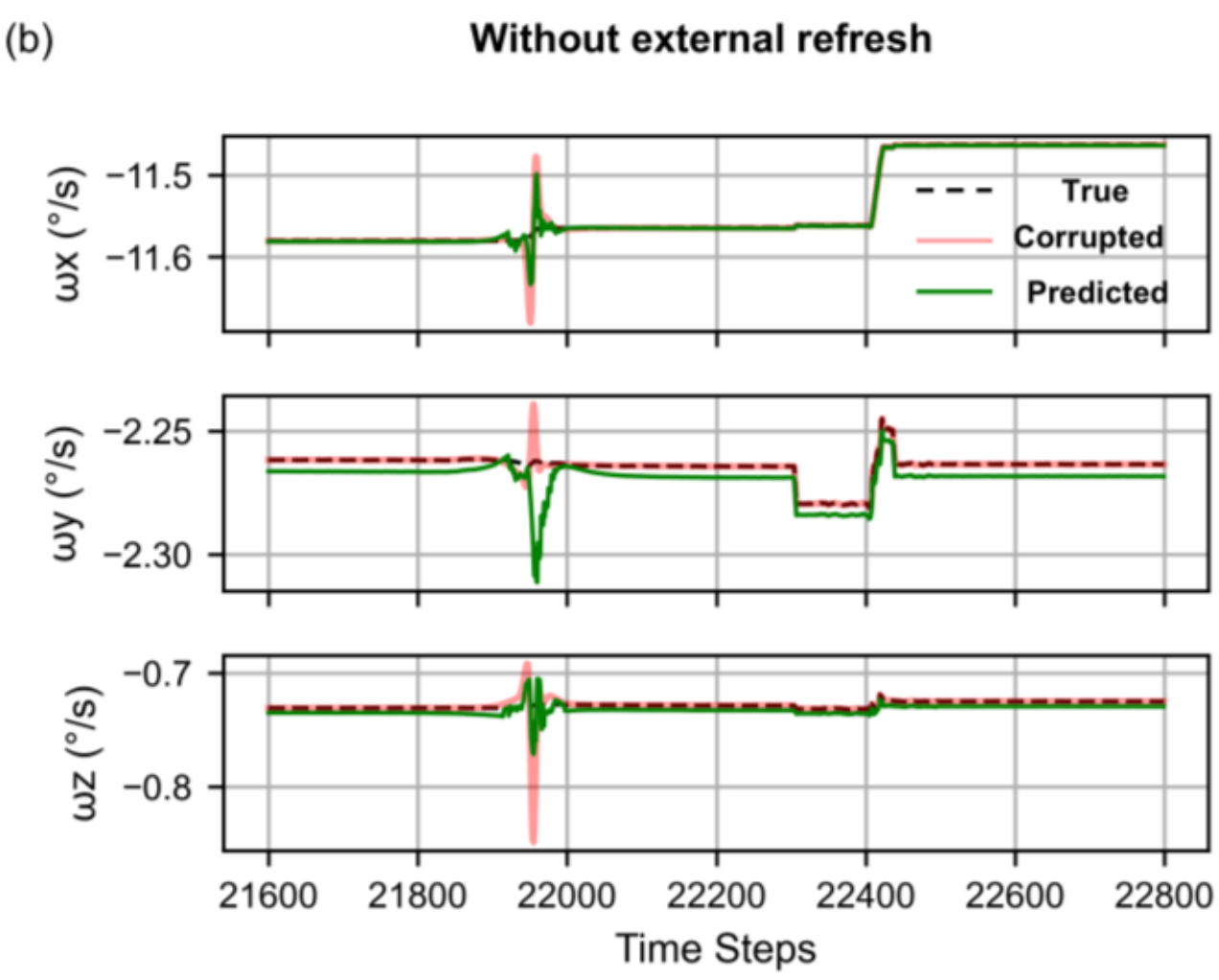

**Figure S15 |** Gyroscope angular velocity post-processing task. (a) The noise reduction rate of gyroscope angular velocity post-processing is 41.46 % with external refresh. (b) The noise reduction rate of gyroscope angular velocity post-processing is 18.88 % without external refresh.

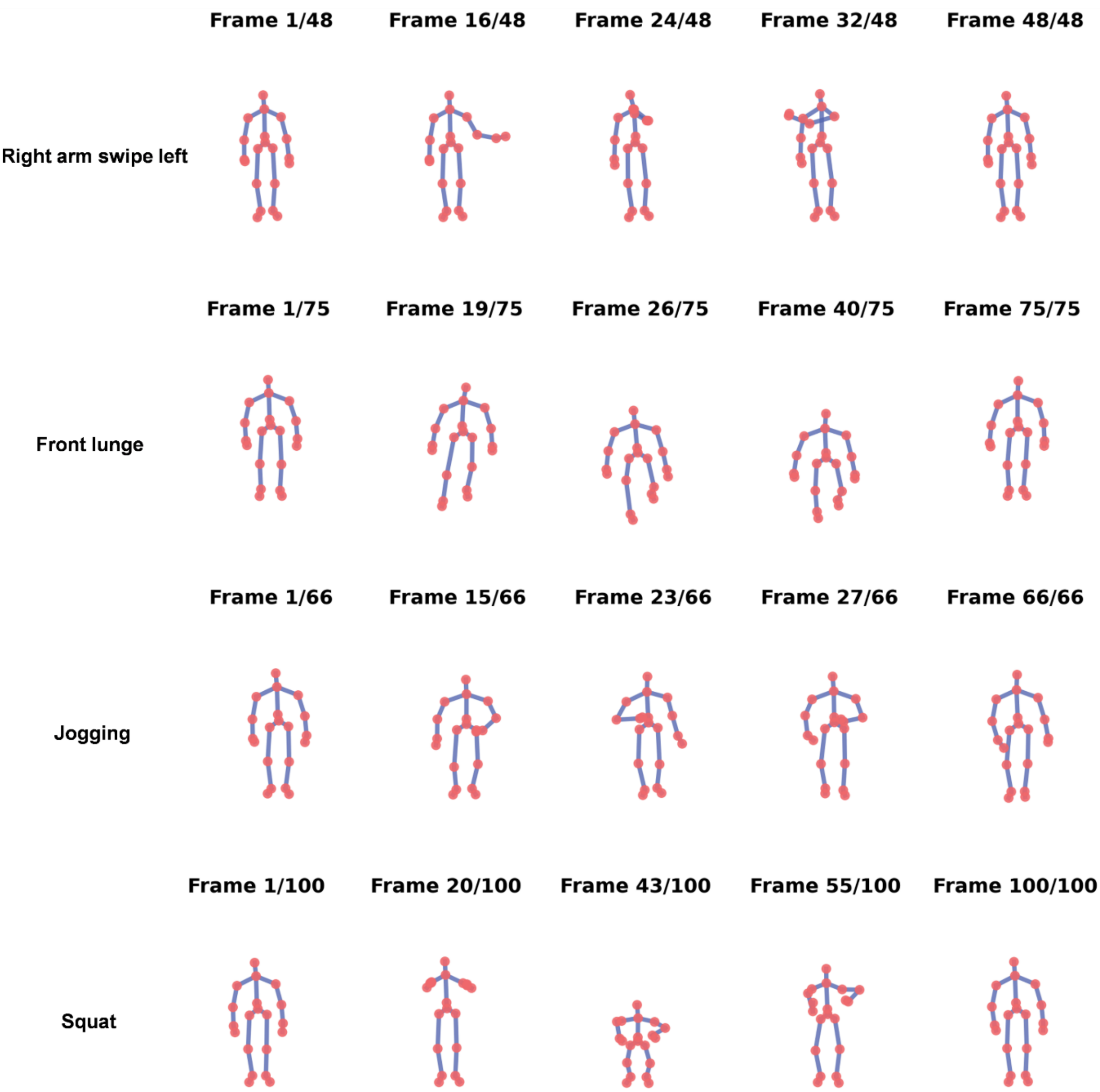

**Figure S16 |** Demonstration of partial UTD-MHAD dataset for action recognition.

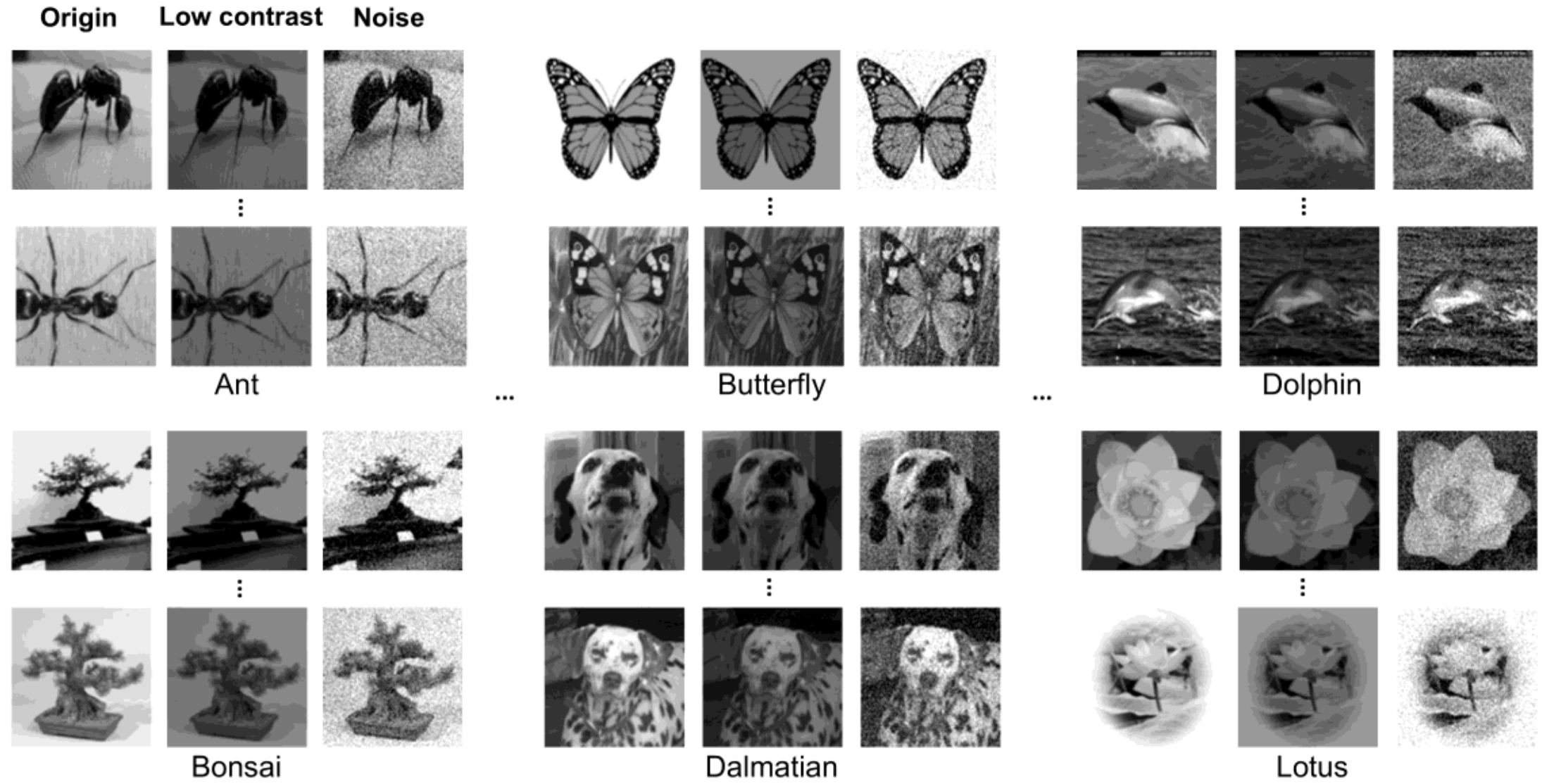

**Figure S17 |** Examples of the 10-class visual object dataset under low-contrast and Gaussian-noise conditions. Representative raw images are adapted from the Caltech-101 database[27].

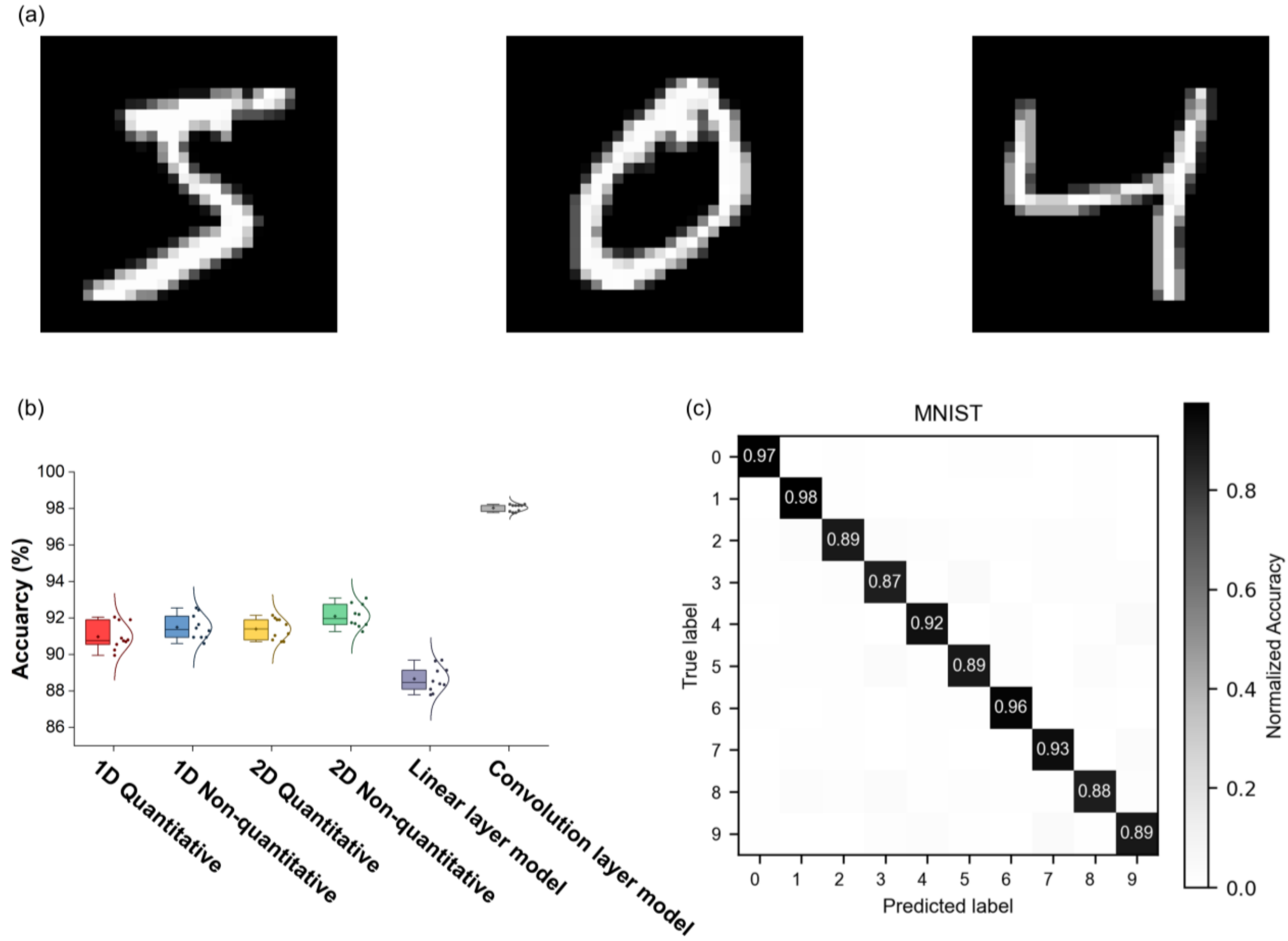

**Figure S18 |** MNIST recognition task[29]. (a) Representative raw images from the MNIST dataset. (b) Comparison of test accuracy across different methods. 1D and 2D denote two distinct coding strategies for pre-processing with our device. Quantitative refers to the classification network implemented on the physical device with discrete conductance states, while Non-quantitative refers to the full-precision software baseline. The two highest test accuracies are 91.4% and 92.11%, achieved respectively by the 2D strategy with quantification and without quantification; both surpass the 88.66% accuracy of a one-layer linear model with full precision. (c) The confusion matrix of the 2D strategy without quantification.

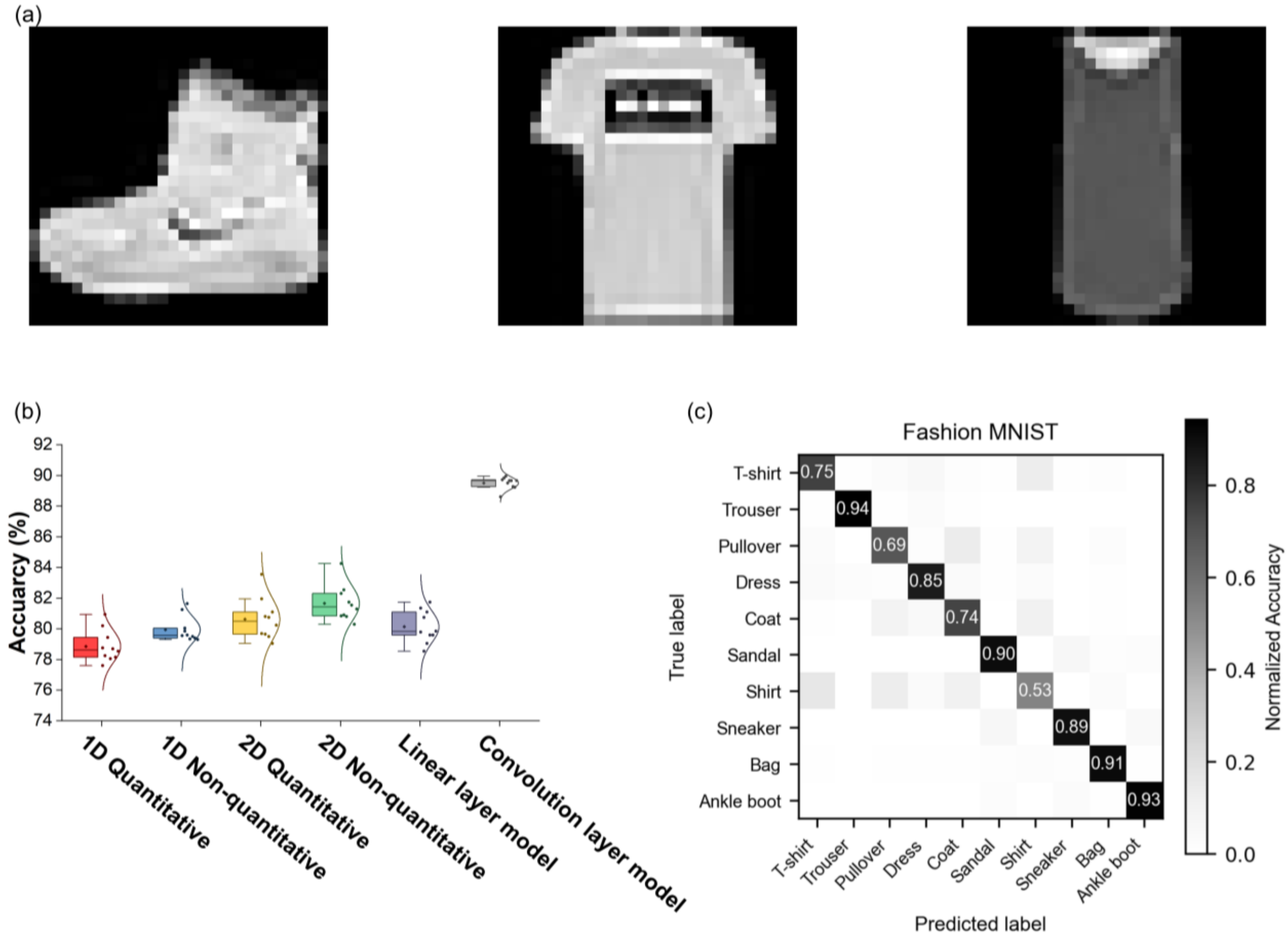

**Figure S19 |** Fashion-MNIST recognition task[30]. (a) Representative raw images from the Fashion-MNIST dataset. (b) Comparison of test accuracy across different methods. The system achieves top accuracies of 80.63% and 81.66% using the 2D strategy with and without quantification, respectively, both of which are higher than the 80.10% achieved by the full-precision linear baseline. (c) The confusion matrix of the 2D strategy without quantification.

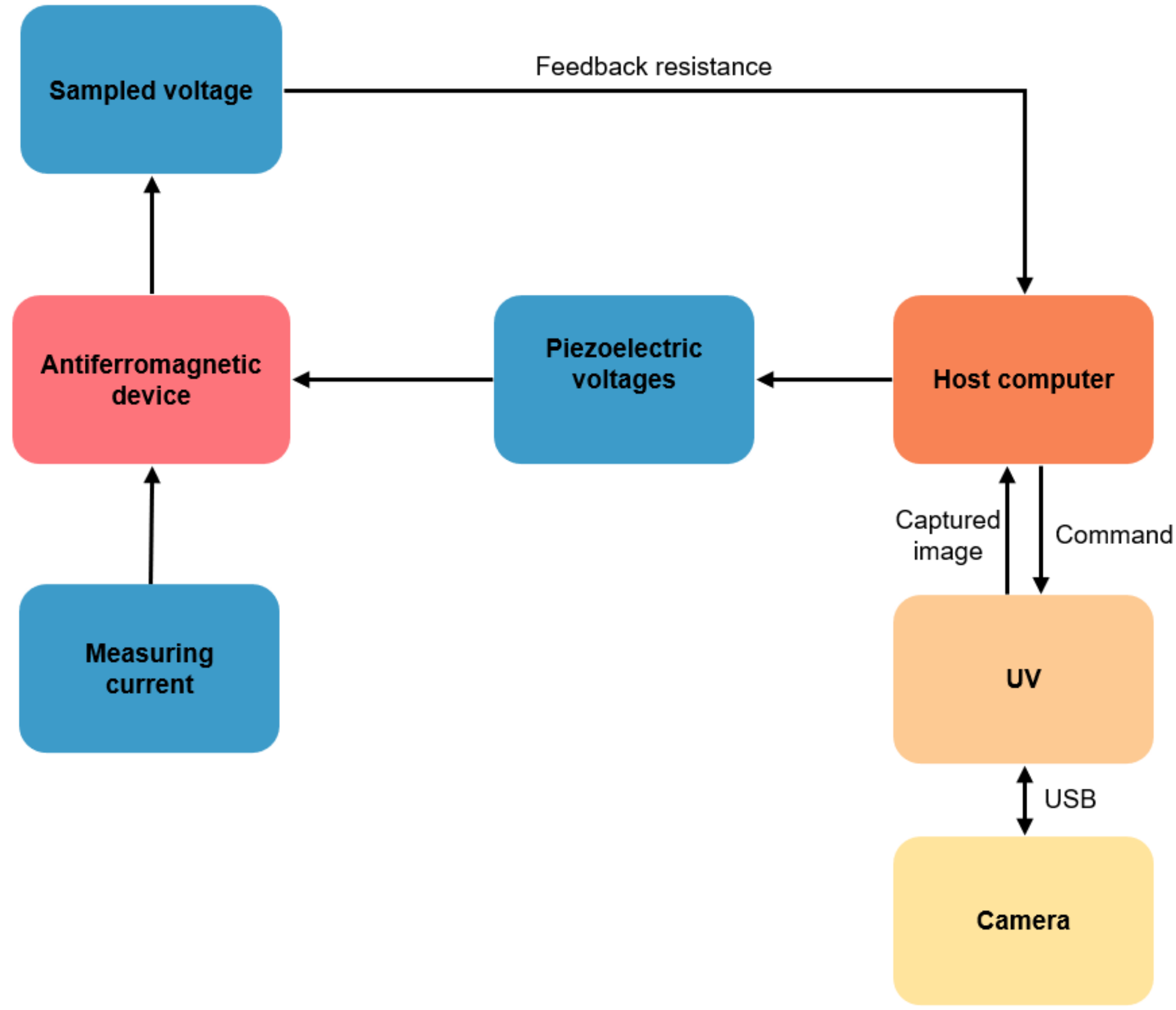

**Figure S20 |** Flow diagram of autonomous UAV system.

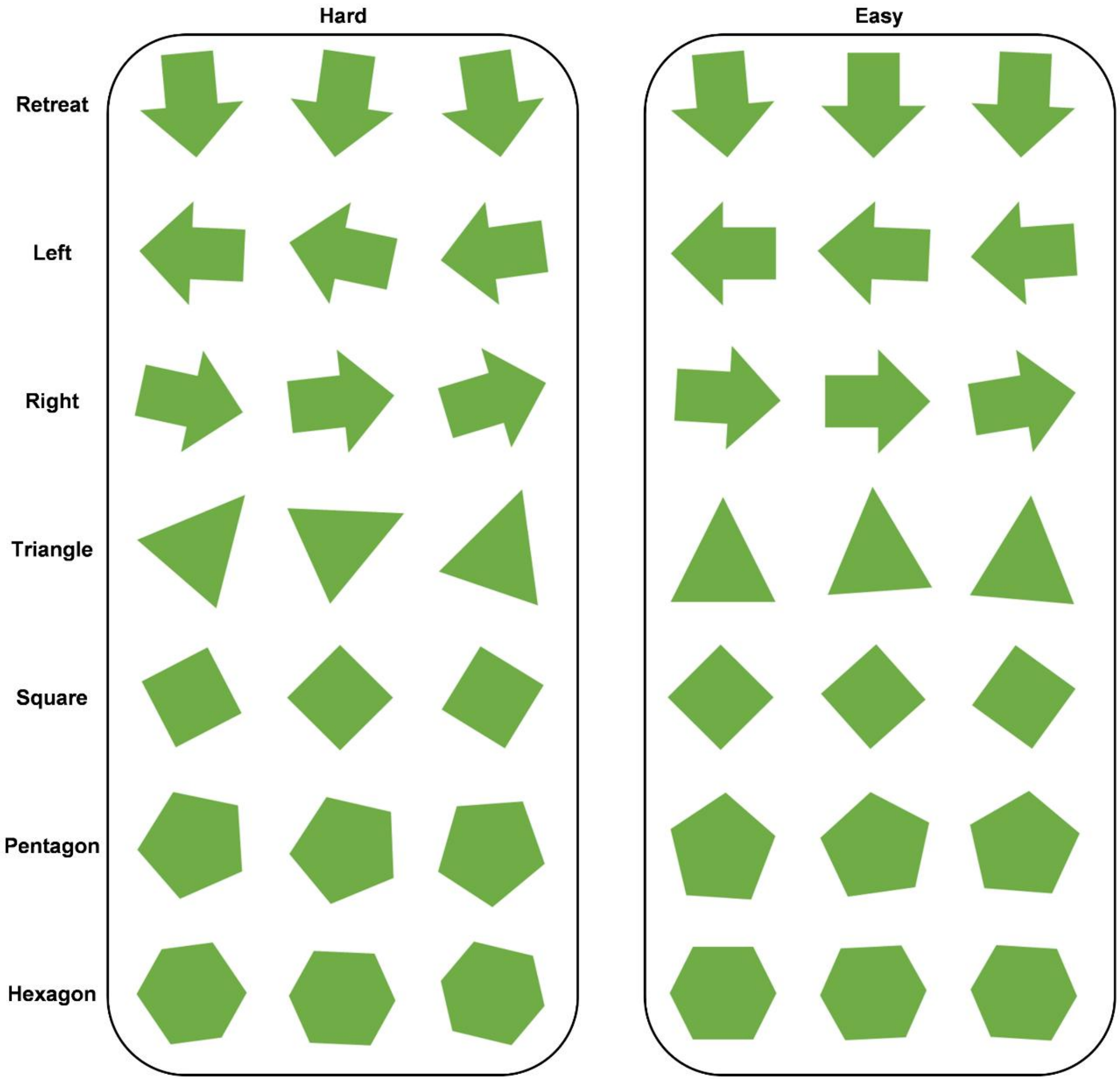

**Figure S21 |** On-board camera dataset for UAV real-time visual recognition.

**Table S1 | Comparison with representative physical computing platforms**

| Physical computing platform | System | Nonlinearity mechanism | Memory mechanism / timescale | State volatility | Drive mode |
|---|---|---|---|---|---|
| Strain-mediated antiferromagnetic | This work (MnIr/PMN-PT) | Strain-mediated AFM domain modulation; harmonic content tunable by minor hysteresis loops (Figure S7) | Nonvolatile domain path dependence; >5-month retention (Figure 2f); fading memory is emulated via self-refreshing | Nonvolatile | Electric field (nA-level leakage) |
| Memristor-based computing | RNPU silicon circuits [14] | Intrinsic nonlinear dynamic transfer functions of memristors | Charge stored by external capacitance (ms scale) | Volatile | Voltage-driven |
| | Dynamic oxide memristor [6] | Rectifying nonlinear *I-V* memristive switching | Dynamic ionic relaxation (ms scale) | Volatile | Voltage-driven |
| | Ag/PVP nanowire network [15] | Ag-filament formation and dissolution in interconnected networks. | Spontaneous dissolution of Ag filaments ($\mu$s–ms scale) | Volatile | Voltage pulses |
| | $WO_X$ memristor [16] | Ion-migration conduction dynamics | Oxygen vacancy back-diffusion ($\mu$s–ms scale) | Volatile | Voltage pulses |
| Spintronic and magnonic computing | Spintronic nano-oscillator [17] | Nonlinear magnetization precession by STT | Magnetization transient relaxation dynamics (ns scale) | Volatile | RF voltage + DC + Magnetic field |
| | Brownian skyrmion reservoir [18] | Voltage-controlled magnetic anisotropy altering the energy landscape of skyrmions | Brownian motion of skyrmions in confined geometry (ms-s scale) | Volatile | Voltage + magnetic field |
| | Magnon-scattering reservoir [19] | Multi-mode magnon interactions | Transient magnon mode lifetimes and cross-stimulation (ns scale) | Volatile | RF/Microwave magnetic field |
| | Magnetic domain Hall-bar [20] | Spatiotemporal transformation in domain-wall | Nonvolatile domain configuration retained after pulse | Nonvolatile | Charge current-induced SOT |
| Ferroelectric and mechanical computing | Ferroelectric diode [21] | Ferroelectric domain nucleation and growth dynamics | Depolarization field-induced partial back-switching (ms scale) | Volatile | Voltage pulse |
| | Ferroelectric MPB transistor [22] | Polarization switching near the mixed phase boundary | Relaxation of HZO near MPB back to non-polarized state ($\mu$s–ms) | Volatile | Voltage pulses |
| | Oxide-based memcapacitor [23] | Ferroelectric polarization switching and charge trapping/detrapping | Depolarization and charge detrapping (ms-s scale) | Volatile | Voltage-driven |
| | MEMS resonator [24] | Mechanical nonlinearity with stiffness modulation | Mechanical ring-down ($\mu$s–ms scale) | Volatile | Stiffness + Electrostatic |

**Table S2 | Summary of Application Scenarios and System Configurations**

| Application scenario | Computational task | Dataset | Signal type | Hardware configuration [a] | Simulated extreme environment | Evaluation metrics |
|---|---|---|---|---|---|---|
| **Space speech recognition** | Time-series classification (10 digits) | TI-46-Word speech dataset [25] | Analog audio waveform | 32 parallel MnIr channels + 2-layer CNN | Baseline benchmarking | 99.8 % Accuracy |
| **Multimodal perception (Action)** | Spatiotemporal skeleton sequence classification (8 classes) | UTD-MHAD [26] | 18-dim skeletal keypoints | 18 MnIr channels + HMnIrSL | Severe solar storm (Radiation and Magnetic interference) | 95.3 % Accuracy |
| **Multimodal perception (Object)** | Static image denoising and classification (10 classes) | Caltech 101 Database (with Gaussian noise & low contrast) [27] | 2D analog pixels | MnIr physical domain filtering + HMnIrSL | Radiation-corrupted imaging conditions | 100 % Accuracy |
| **Multimodal perception (Gesture)** | Real-time dynamic posture classification (5 classes) | Real-time MPU-6050 Accelerometer | 3-axis acceleration (Time-series) | MnIr feature mapping + HMnIrSL | Severe solar storm | 98.0 % Accuracy |
| **Spacecraft signal processing** | Time-series denoising | NASA Juno probe flight data (PJ45-PJ63) | 3D angular velocity + 3D magnetic field | 6-dim inputs to parallel MnIr array with external refresh + Ridge regression | Jovian magnetosphere | 41.46 % Noise reduction rate (RMSE ~ $10^{-3}$ °/s) |
| **Autonomous drone navigation** | Real-time visual recognition and command generation | On-board camera video stream | Dynamic 2D patches (10×10 pixels) | Parallel spatiotemporal coding by MnIr + HMnIrSL | GPS-denied conditions | ~100 % (Easy) / 82 % (Hard) Accuracy |

[a] Hardware-constrained MnIr Synaptic layer (HMnIrSL)